\documentclass[prd, nofootinbib, notitlepage]{revtex4-1}

\usepackage{graphicx}
\usepackage{caption}
\usepackage[labelformat=simple]{subcaption}

\usepackage{amsmath,bm}
\usepackage[T1]{fontenc}
\usepackage{rotating}
\usepackage{float}
\floatstyle{plaintop}
\restylefloat{table}

\usepackage{color}
\definecolor{naviBlue}{RGB}{0,0,128}
\usepackage[colorlinks  = true,
            linkcolor   = naviBlue,
            urlcolor    = naviBlue,
            citecolor   = naviBlue,
            anchorcolor = naviBlue]{hyperref}
\newcommand{\secref}[1]{\hyperref[sec::#1]{SECTION~\ref*{sec::#1}}}
\newcommand{\subsecref}[1]{\hyperref[subsec::#1]{SECTION.~\ref*{subsec::#1}}}
\newcommand{\figref}[1]{\hyperref[fig::#1]{FIG.$\,$\ref*{fig::#1}}}
\newcommand{\tabref}[1]{\hyperref[tab::#1]{TABLE$\,$\ref*{tab::#1}}}
\newcommand{\eqnref}[1]{\hyperref[eqn::#1]{Eq.$\,$(\ref*{eqn::#1})}}

\newcommand{\diff}{\mathrm{d}}
\newcommand{\I}[1]{\textit{#1}}

\newcommand{\ac}[1]{#1}

\begin{document}

\title{Galactic cosmic-ray propagation in the light of AMS-02: I. Analysis of protons, helium, and antiprotons}

\author{Michael Korsmeier}
\email{korsmeier@physik.rwth-aachen.de}
\author{Alessandro Cuoco}
\email{cuoco@physik.rwth-aachen.de}
\affiliation{Institute for Theoretical Particle Physics and Cosmology, RWTH Aachen University, 52056 Aachen, Germany}

\begin{abstract}
We present novel constraints on cosmic-ray propagation
in the Galaxy using the recent precise measurements of proton and helium
spectra from AMS-02, together with preliminary AMS-02 data on the antiproton over proton ratio.
To explore efficiently the large (up to eleven-dimensional) parameter space
we employ the nested-sampling algorithm as implemented in the \textsc{MultiNest} package,
interfaced with the \textsc{Galprop} code to compute the model-predicted spectra.
We use VOYAGER proton and helium data, sampling the local interstellar spectra,
to constrain the solar modulation potential.
We find that the turbulence of the Galactic magnetic field is
well constrained,  i.e., $\delta=0.30^{+0.03}_{-0.02}(stat)^{+0.10}_{-0.04}(sys)$, with uncertainties dominated
by systematic effects.  Systematic uncertainties
are determined checking the robustness of the results to the minimum
rigidity cut used to fit the data (from 1$\,$GV to 5$\,$GV), 
to the propagation scenario (convection vs no convection),
and to the uncertainties
in the knowledge of the antiproton production cross section.
Convection and reacceleration are found to be degenerate and not well constrained
singularly when using data above 5$\,$GV.  Using data above 1$\,$GV  reacceleration is required, $v_{\rm A}=25\pm2$km/s,
 although this value might be significantly affected by the low-energy systematic uncertainty in the solar modulation.
In a forthcoming companion paper, we investigate the constraints imposed by AMS-02
measurements on  lithium, boron, and carbon.

\end{abstract}

\maketitle



\section*{\label{sec::introduction}Introduction}

Cosmic-ray (CR) physics is  on the verge of transition to a precision era   thanks to the recently available data from PAMELA
first, and more recently from the AMS-02 experiment on board the International Space Station.        
Thanks to these precise data, cracks in the standard minimal scenario start to appear.
For example, a significant difference in the slopes of proton and helium, 
of about $\sim$0.1  \cite{Adriani_PAMELA_pHe_2011,Aguilar_AMS_Proton_2015,Aguilar_AMS_Helium_2015}), has been observed,
while, from the standard CR acceleration scenario, no differences would be expected, 
at least for energies above 20-30$\,$GeVs.
The same measurements also find a break in the proton and helium rigidity spectra at about 300$\,$GV.
In this case, the feature can be accommodated with an extension of the standard scenario,
and various explanations have been proposed \cite{Vladimirov:2011rn,Serpico:2015caa,Blasi:2012yr,Blasi:2011fi,Aloisio:2015rsa}.

Nonetheless, besides the above CR `anomalies', the standard \emph{diffusion-reacceleration-convection} scenario 
is, in the first place, not yet very well constrained.  For example, estimates of the degree of turbulence in the Galactic magnetic field 
(encoded in the parameter conventionally indicated as $\delta$) range from the standard Kolmogorov
turbulence ($\delta=0.33$, \cite{Trotta_CR_Propagation_2011,Johannesson_CR_Propagation_2016})
to Kraichnan ($\delta=0.5$) or plain diffusion ($\delta=0.6$, \cite{Bernardo_Unified_CR_interpretation_2010}),
up to $\delta=0.9$ \cite{Maurin:2001sj,Putze_MCMC_CR_BoverC_2010}.
Again, the new precise data offer the possibility to finally pin down the uncertainties in
the parameters of the model.

In the following, we will thus analyze recently published proton  \cite{Aguilar_AMS_Proton_2015},
and helium  \cite{Aguilar_AMS_Helium_2015} AMS-02 data together with preliminary AMS-02
data  on the antiproton over proton ratio \cite{Kounine_AMS_pbar_2015}.
The analysis of the heavier nuclei, lithium, boron and carbon and comparison
with the results from the analysis of this work is presented in a companion forthcoming article.

We also treat in a novel way the effect of solar modulation.
Although we still use the force-field approximation, we do not assume any prior on
the solar modulation potential, but we, instead, use recent VOYAGER data \cite{Stone_VOYAGER_CR_LIS_FLUX_2013},
sampling the interstellar unmodulated CR spectrum, to constrain the amount
of solar modulation.  
The effect of solar modulation will be also studied applying different cuts on the minimum
rigidity of the data used in the fit.
Finally, we will also investigate the effect of uncertainties in the antiproton production cross section,
following the recent redetermination from \cite{Mauro_Antiproton_Cross_Section_2014}.

\ac{There is another well-known anomaly in CRs, namely the rising positron fraction observed by both PAMELA \cite{Adriani:2008zr} and AMS-02 \cite{Aguilar:2013qda}. The rising is incompatible with the usual interpretation of positrons as secondaries produced by protons during their propagation.
Although some attempt has been made to reconcile the positron fraction with the interpretation as secondaries through some modification of the propagation model \cite{Shaviv:2009bu}, the generally accepted  explanation requires a primary source of positrons, like pulsars or a nearby supernova remnant, or possibly, dark matter annihilation.
In our study, we will assume the standard propagation scenario described above, and, as such, we would require a primary positron source  
to explain the observations.
Nonetheless, even after including a primary source, it might be nontrivial to explain the positron fraction as well as the $e^++e^-$ spectrum,
since the propagation of leptons is significantly affected by energy losses in the local radiation and magnetic field,
while the local turbulence properties of the magnetic field can be different from the large scale average probed by nuclei.
Although it would be, thus, interesting to cross-check the results of the study of propagation of nuclei 
with lepton spectra observations, ultimately, the above issues would make the comparison complicated
and difficult to interpret.   
We will thus avoid these comparisons in the following and focus only on nuclei.
}

The work is structured as follows:
The theoretical framework is discussed in \secref{theory}. 
The fit methodology is discussed in \secref{methods}.
The results are presented in \secref{results}, while we conclude in \secref{conclusion}.

\section{\label{sec::theory}Theory}

The propagation of CR can be described by the well-known diffusion equation \cite{StrongMoskalenko_CR_rewview_2007} for the 
particle density $\psi_i$ of species $i$ per volume and absolute value of momentum $p$
\begin{eqnarray}
  \label{eqn::PropagationEquation}
  \frac{\partial \psi_i (\bm{x}, p, t)}{\partial t} = 
    q_i(\bm{x}, p) &+&  
    \bm{\nabla} \cdot \left(  D_{xx} \bm{\nabla} \psi_i - \bm{V} \psi_i \right) \nonumber \\ 
     &+&  \frac{\partial}{\partial p} p^2 D_{pp} \frac{\partial}{\partial p} \frac{1}{p^2} \psi_i - 
    \frac{\partial}{\partial p} \left( \frac{\diff p}{\diff t} \psi_i  - \frac{p}{3} (\bm{\nabla \cdot V}) \psi_i \right) -
    \frac{1}{\tau_{f,i}} \psi_i - \frac{1}{\tau_{r,i}} \psi_i.
\end{eqnarray}

The various terms describe (i) spatial diffusion, usually assumed to be homogeneous and isotropic and thus described by the momentum-dependent
diffusion coefficient $D_{xx}(p)$,  (ii) convective winds, described by their velocity $\bm{V}(\bm{x})$,  
(iii) diffusive reacceleration, parametrized as a diffusion in momentum space with coefficient  $D_{pp}(p)$,
(iv) continuous energy losses through the coefficient $dp/dt=\sum_k dp_k/dt$ which sums over all the various processes, $dp_k/dt$, 
through which the particles lose energy, (v) adiabatic energy losses, present if $\bm{V}(\bm{x})$ has a nonzero divergence,
and finally, catastrophic losses by (vi) decay or (vii) fragmentation, with  decay and interaction times $\tau_r$ and $\tau_f$, respectively.
The equation is typically solved assuming a steady state regime, meaning that $\psi_i$ does not depend on time and so the term on left-hand side is zero.

Diffusion is naturally expected to be an energy dependent process,
with particles being less deflected by the  magnetic fields with  increasing energy, and thus diffusing faster. 
This process is usually modeled by a power law in rigidity $R=p/|Z|$ (\cite{Blandford:1987pw}):
\begin{eqnarray}
  \label{eqn::DiffusionConstant}
  D_{xx} &= \beta D_{0} \left( \frac{R}{4 \, \mathrm{GV}} \right)^{\delta},
\end{eqnarray}
where $\delta$ is the index of the power-law, $D_0$ the overall normalization, and
$\beta=v/c$ the velocity of the CRs; 
we set the normalization scale at 4$\,$GV.
The constant for diffusive reacceleration $D_{pp}$ is usually related to the spatial diffusion $D_{xx}$ and 
to the velocity $v_\mathrm{A}$ of Alfven magnetic waves \cite{Ginzburg:1990sk,1994ApJ...431..705S} as
\begin{eqnarray}
  \label{eqn::DiffusivReaccelerationConstant}
  D_{pp} = \frac{4 \left(p \, v_\mathrm{A} \right)^2 }{3(2-\delta)(2+\delta)(4-\delta)\, \delta \, D_{xx}}.
\end{eqnarray}
The amount of reacceleration is thus described in terms of the parameter $v_\mathrm{A}$.
Finally, convective winds are assumed to be constant and orthogonal to the Galactic plane  $\bm{V}(\bm{x})= {\rm sign}(z)\, v_{0,c} $.
We note that, in principle, this parametrization  implies an unphysical discontinuity
at $z=0$. A smooth transition in the thin halo containing the sources (with size $\sim $0.2 kpc) would be more realistic.
Nonetheless, since this parametrization has been widely employed in past works, we use it for the sake of comparison.

The source term $q_i(\bm{x}, p)$ of primary CR is assumed to factorize into a species dependent normalization $q_{0,i}$, 
a space-depend part $q_{r,z}$ (where $r=\sqrt{x^2+y^2}$ and $z$ are Galactocentric cylindrical coordinates), 
and  a rigidity dependent part $q_R$:
\begin{eqnarray}
  \label{eqn::SourceTerm_1}
  q_i(\bm{x}, p) = q_i(r, z, R) = q_{0,i} \ q_{r,z}(r,z) \, q_R(R).
\end{eqnarray}
We model the rigidity dependence as double  broken power law  with smooth transitions
\begin{eqnarray}
  \label{eqn::SourceTerm_2}
  q_R(R)     &=&   \left( \frac{R}{R_0} \right)^{-\gamma_1}
                  \left( \frac{R_0^{\frac{1}{s}}+R^  {\frac{1}{s}}}
                              {2(R_0)^{\frac{1}{s}}                } \right)^{-s (\gamma_2-\gamma_1)}
                  \left( \frac{R_1^{\frac{1}{s_1}}+R^  {\frac{1}{s_1}}}
                              { R_1^{\frac{1}{s_1}}                   } \right)^{-s_1(\gamma_3 - \gamma_2)},
\end{eqnarray}
where $R_0$, $R_1$ are the two break positions, $s$, $s_1$ the smoothing factors, and $\gamma_i$ ($i=1,2,3$) the slopes
in the various rigidity ranges in between the breaks. The normalization is such that  $q_R(R) =1$ at $R=R_0$. 
Typically, only one break has been  considered in the literature, 
with value of the order $\sim 10\,$GV \cite{Trotta_CR_Propagation_2011},
or none\footnote{
In \cite{Putze_MCMC_CR_LeakyBox_2009,Putze_MCMC_CR_BoverC_2010}
a source term $q \propto \beta^{-1} R^{-\gamma}$ is considered,
which implies a break in momentum at a rigidity $\sim m/Z$, with an upward
steepening of 1 in the slope.}  \cite{Bernardo_Unified_CR_interpretation_2010}.
On the other hand the recent discovery of a break at around
$300\,$GV in the proton and helium spectra first by PAMELA \cite{Adriani_PAMELA_pHe_2011} and then by AMS-02 \cite{Aguilar_AMS_Proton_2015,Aguilar_AMS_Helium_2015} makes it necessary
to introduce a second break for a proper description of the data.
This was, indeed, considered, for example, in  \cite{Johannesson_CR_Propagation_2016,Evoli:2015vaa}.
We further introduced in \eqnref{SourceTerm_2}, as a novel feature with respect to previous studies,
the parameters $s_i$ to explore the possibility of a smooth transition between the various  regimes, as opposed to a sharp one.

We mention here that an alternative possibility would be to model the break as a break in the diffusion
rather than the injection spectrum.
This has the same effect for the primaries' spectra but leads to different results for secondaries. 
The secondaries' injection spectra would reflect the break from the primary spectra, but the amount of the break would increase during the
propagation, with the result that  the break is expected to be twice as large as the one of primaries.
Nonetheless, for antiprotons this effect would start to be significant only at very large energies (above few hundreds GV),
which are not yet well measured by AMS-02, and thus the two scenarios are equivalent.
The effect could be, instead, important for lithium or boron  AMS-02 measurements, which extend to larger
energies with respect to antiprotons.

The spatial dependence, i.e., the source distribution, is parametrized as 
\begin{eqnarray}
  \label{eqn::SourceTerm_3}
  q_{r,z}(r, z)  &= \left( \frac{r}{r_s} \right)^\alpha \exp \left( -\beta \frac{r-r_s}{r_s} \right) 
                                                    \exp \left( -      \frac{|z|  }{z_0} \right),
\end{eqnarray}
with parameters $\alpha = 0.5$, $\beta=1.0$, $r_s=8.5$ kpc, and $z_0=0.2$ kpc. For the analysis of $\gamma$ rays one usually uses source distribution inferred from pulsars \cite{Yusifov_Pulsar_Distribution_2004} or supernova remnants \cite{Case_SNR_Distribution_1998,Green:2015isa}. Typical parameter values in those cases are  $\alpha \sim 1.6$, $\beta \sim 4$ with a flattening above $r\gtrsim10\,$kpc and a cutoff above $r\gtrsim30\,$kpc. We checked that changing the source distribution to those values has a negligible impact on the CR  energy spectra after propagation.

In the case of secondary CRs, as for antiprotons produced by primary CRs through spallation in the 
interstellar medium (ISM), the source term is given by the primaries themselves.
More precisely the source term is the integral over the momentum-dependent production rate of the secondaries and the sum 
over the primary species $i$ and the ISM components $j$,
\begin{eqnarray}
  \label{eqn::SourceTerm_pbar}
  q(\bm{x},p) = \sum\limits_{j=\mathrm{H,He}} n_j(\bm{x}) \sum\limits_{i=\mathrm{p,He}} 
    \int \diff p_i \, \frac{\diff \sigma_{ij}(p, p_i)}{\diff p} \beta_i \, c\, \psi_i(\bm{x},p_i),
\end{eqnarray}
where $\sigma_{ij}$ is the antiproton production cross section by the species $i$ spallating over the ISM species $j$.
The ISM is assumed to be composed of hydrogen and helium gas with fixed  proportion 1:0.11.
The abundance of secondaries is typically quite low with respect to the primaries, and
this allows one to evaluate \eqnref{SourceTerm_pbar} with $\psi_j(\bm{x},p_i)$ calculated
from \eqnref{PropagationEquation} neglecting in the first place the secondaries.  
\ac{Besides antiprotons, we will consider also secondary protons, 
i.e., primary protons that underwent inelastic scattering, losing a substantial fraction of their
energy, and thus reappearing at low energies.
We will also take into account tertiary antiprotons produced by the spallation of the secondary 
antiprotons during propagation.   Secondary protons and tertiary antiprotons are described with the same formalism. 
Their source term can be calculated analogously to \eqnref{SourceTerm_pbar}
but replacing $\psi_i(\bm{x},p_i)$ with the density of primary   protons in the first case, and secondary antiprotons in the second case,
and using the associated production cross section.
The latter is approximated as the total inelastic non annihilating cross section of the incoming proton or antiproton times
the energy distribution of the scattered particle, approximated as $1/E_{\rm kin}$.
For more details see Ref.~\cite{Moskalenko:2001ya}.}

To numerically solve the propagation equation \eqnref{PropagationEquation} 
and to derive the secondaries' and tertiaries' abundances we use
the  \textsc{Galprop} code\footnote{http://galprop.stanford.edu/} \cite{Strong:1998fr,Strong:2015zva}.
We use version $r2766$\footnote{https://sourceforge.net/projects/galprop/} as basis, and we implement some custom modifications,
such as the possibility to use species-dependent injection spectra, which is not allowed by default in \textsc{Galprop}. Furthermore, we allow for a smoothing of the originally simple broken power law as discussed above. 

The propagation equation \eqnref{PropagationEquation} is solved on a grid  in the energy dimension and in the two spatial dimensions $r$ and $z$, assuming cylindrical symmetry of our Galaxy.
The radial boundary of the Galaxy is fixed to $20\,$kpc, while the half-height $z_h$
 is a free parameter. The radial and $z$ grid steps are chosen as $\Delta r=1\,$kpc, and $\Delta z = 0.2\,$kpc. 
The grid in kinetic energy per nucleon is logarithmic between $1$ and $10^7\,$MeV with a step factor of $1.4$. 
Free escape boundary conditions are used, imposing  $\psi_i$ equal to zero outside the region sampled by the grid. 
We tested also more accurate choices for the above settings and found the results stable against the changes.

\ac{Note also that we consider propagation of nuclei only up to $Z$=2,  i.e., in practice, in  \textsc{Galprop}  
we propagate  $p$, $\bar{p}$, $^2$H, $^3$He, and $^4$He  species plus the secondary protons  and the tertiaries antiprotons.
This also means that we neglect possible contributions from the fragmentation of  $Z>$2 nuclei,
which should be a good approximation since their fluxes are much lower than the $p$ and He fluxes.
Nonetheless, in the specific case of our best-fit propagation scenario (see below),   we verified explicitly that
including nuclei with $Z>$2 in the calculation changes the spectra of He (i.e., $^3$He + $^4$He) only by few percent
and protons (i.e., $p$ + $^2$H)  by less than 1\%.  
This is also confirmed by the study in ref.\cite{Coste:2011jc}, where it is also shown that the $Z>$2 nuclei contribution to He is few \%
(although the  contribution to $^2$H, $^3$He can be, instead, up to 20-30\%).}

\section{\label{sec::methods}Methods and Data}

\begin{table}[b!]
  \caption{Summary of the data-sets used in this analysis. }
  \label{tab::FitData}
  \centering
  \begin{tabular}{l l r c l l}
  \hline \hline
  \textbf{Experiment} & \textbf{Species} & \multicolumn{3}{l}{\textbf{Rigidity range [GV]} } &  \textbf{ref.} \\ \hline \hline
  AMS-02   & Proton $           $ & $1.0\cdot 10^0$ & - & $1.8\cdot 10^3$ & \cite{Aguilar_AMS_Proton_2015} \\
  AMS-02   & Helium $           $ & $1.9\cdot 10^0$ & - & $3.0\cdot 10^3$ & \cite{Aguilar_AMS_Helium_2015} \\
  AMS-02   & Antiproton ratio $ $ & $1.0\cdot 10^0$ & - & $0.2\cdot 10^3$ & \cite{Kounine_AMS_pbar_2015} \\
  CREAM    & Proton $           $ & $3.2\cdot 10^3$ & - & $2.0\cdot 10^5$ & \cite{Yoon_CREAM_CR_ProtonHelium_2011} \\
  CREAM    & Helium $           $ & $1.6\cdot 10^3$ & - & $1.0\cdot 10^5$ & \cite{Yoon_CREAM_CR_ProtonHelium_2011} \\
  VOYAGER  & Proton $           $ & $0.7\cdot 10^0$ & - & $1.0\cdot 10^0$ & \cite{Stone_VOYAGER_CR_LIS_FLUX_2013} \\
  VOYAGER  & Helium $           $ & $0.6\cdot 10^0$ & - & $2.3\cdot 10^0$ & \cite{Stone_VOYAGER_CR_LIS_FLUX_2013} \\ \hline \hline
  \end{tabular}
\end{table}

\subsection{Data}

As described in the introduction, the main focus of the analysis is on the new AMS-02 measurements.
We will thus use the published proton \cite{Aguilar_AMS_Proton_2015} and helium \cite{Aguilar_AMS_Helium_2015} 
AMS-02 spectra, and the available preliminary measurements of the antiproton over proton ratio \cite{Kounine_AMS_pbar_2015}.
The AMS-02 $p$ and He data extend up to a few TV. We thus complement them with $p$ and He CREAM measurements
starting from a few TV up to $\sim$ 100 TV.
Finally, we use recently measured $p$ and He VOYAGER data  \cite{Stone_VOYAGER_CR_LIS_FLUX_2013} 
at low rigidities $\alt$ $1\,$GV  which are believed to be the first
direct measurement of the
local interstellar (LIS) flux, as a consequence of the fact that the  probe crossed the solar Helio pause,
leaving the solar system and entering in the interstellar space.
A summary of the  data sets  used is presented in \tabref{FitData}.
We use all data in rigidity, since this is the directly measured quantity by AMS-02 as opposed to the kinetic energy.

\subsection{Solar modulation}

To compare the AMS-02 fluxes to the \textsc{Galprop} model predictions solar effects have to be taken into account. 
CRs are deflected and decelerated in the solar winds, whose activity varies in a 22 year cycle. 
The effect of this solar modulation \cite{Parker_SolarModulation_1958}  can be described phenomenologically by the 
force-field approximation \cite{Usoskin_Solar_Modulation_2005,Gleeson_SolarModulation_1968}, 
which is equivalent to taking into account only the adiabatic energy losses of the CRs
propagating in the expanding solar wind.
The process can be described by a single parameter $\phi$, the solar modulation or Fisk potential \cite{Fisk_SolarModulation_1976},
which links  the total energy of the particles in the local interstellar space $E_\mathrm{LIS}$  to the energy $E$ observed
in the detector at the Earth.
The energy-differential flux $\Phi_E$ is then modulated by
\begin{eqnarray}
  \label{eqn::SolarModulation_ForceFieldApproximation}
  E &=& E_\mathrm{LIS}- |Z| e \phi \\
  \Phi_E(E) &=& \frac{E^2-m^2}{E_\mathrm{LIS}^2-m^2} \Phi_{E,\mathrm{LIS}}(E_\mathrm{LIS}),
\end{eqnarray}
where $Z$ is the charge number, $e$ the elementary charge, and $m$ the mass. 
The modulation potential can be approximately derived from measurements of the neutron flux
at Earth by various neutron monitor stations, since a strong anticorrelation is observed between
the neutron flux and the solar activity \cite{Usoskin_Solar_Modulation_2005, Usoskin_Solar_Modulation_2011}. Nonetheless, the procedure
is affected by large uncertainties.
In previous works, the usual  procedure was  to  use this value, associate with it a ``reasonable''
uncertainty,  and use it as a prior in the fit to CR data.
Here, instead, we will use a novel procedure, similar to the one implemented in  
\cite{Cholis_Solar_Modulation_2016}. Assuming the measured VOYAGER $p$ and He fluxes,
indeed, sample the LIS fluxes, we fit them with the unmodulated spectra,
 while, at the same time,  the modulated spectra are  fitted to the AMS-02 data.
CREAM data, instead, are at very high energies where solar effects can be neglected.
The explicitly used $\chi^2$ is reported in the next section in \eqnref{ChiSquare}.
We thus do not assume any specific prior for $\phi$ (in practice allowing a very large range, see Table 2),
and we let the VOYAGER data constrain it.
In the future, monthly or weekly $p$,  He, and $\bar{p}$ data from AMS-02 should further help in better
constraining $\phi$.

The  force-field approximation, nonetheless, gives only a first-order description of the solar modulation process.
A more complete description relies on a transport equation analogous to \eqnref{PropagationEquation} but including the
specific processes experienced by CRs while propagating in the solar magnetosphere \cite{Potgieter:2013pdj}.
The implementation of these models is, however, beyond the scope of the present work.
Indeed, dedicated analyses, using time dependent proton flux of PAMELA \cite{Adriani:2013as,Potgieter:2013cwj}
 and VOYAGER data,
  suggest a strong rigidity and charge sign dependency of $\phi$ below a rigidity $5\,$GV \cite{Cholis_Solar_Modulation_2016,Corti:2015bqi},
  indicating, in other words, a breakdown of the force-field approximation.
Therefore, in the present analysis we use a fiducial lower rigidity threshold of $5\,$GV,
although,  we will also compare our fiducial fit results with those obtained including data down to $1\,$GV.

As a final comment, we  note that while \eqnref{SolarModulation_ForceFieldApproximation} is non linear,
only a single average potential $\phi$ is used in it. On the other hand, $\phi$ typically undergoes significant variations
during the entire period over which the final averaged measured spectrum is provided  
(\ac{see \cite{Ghelfi:2016pcv} for a recent study of the time variation of $\phi$}).
We thus tested a fictitious case in which $\phi$ varies linearly in time  from a value of 300$\,$MV to 700$\,$MV
during a period of 2 years and we 
applied to a given model LIS spectrum the force-field approximation in small time bins, using in each bin
the appropriate potential and then averaging at the end to derive the final modulated spectrum.
This was compared with the flux obtained by direct application of \eqnref{SolarModulation_ForceFieldApproximation}   
to the LIS using the average $\phi$.  We found no appreciable difference between the two,
indicating that \eqnref{SolarModulation_ForceFieldApproximation} behaves linearly to a very good
approximation. A posteriori, this can be justified in terms of the smallness of the $\phi$ parameter with respect
to the larger rigidities involved.

\subsection{Fit procedure}

To scan the large parameter space we use the \textsc{MultiNest} package \cite{Feroz_MultiNest_2008}. 
\textsc{MultiNest} implements the algorithm of ellipsoidal nested sampling \cite{Skilling_NestedSampling_2006}, 
allowing efficient likelihood evaluation and evidence calculation. 
As likelihood we use $\mathcal{L} = \exp\left( -\chi^2/2 \right)$ with  
\begin{eqnarray}
  \label{eqn::ChiSquare}
  \chi^2 &=  \sum\limits_i \frac{(\Phi_{AMS,i} - \Phi_{M}(R_i))^2}{\sigma_{AMS,i}^2}   +
    \sum\limits_{D=V, C} \sum\limits_i \frac{(\Phi_{D,i} - \Phi_{M,LIS}(R_i))^2}{\sigma_{D,i}^2}     ,
\end{eqnarray}
where $D=V,C$ indicates the VOYAGER and CREAM data sets with data points $i$ at rigidity and flux ($R_i$, $\Phi_{D,i}$), 
with measured flux uncertainties  flux $\sigma_{D,i}$, and 
$\Phi_{M,LIS}$  is the unmodulated model differential flux.
Analogously,  $\Phi_{AMS,i}$ and $\sigma_{AMS,i}$  indicates the AMS-02  data points and uncertainties,
and  $\Phi_{M}$ is the model flux modulated with a potential appropriate for AMS-02, $\phi_{\rm AMS}$. 
\ac{We note here that reported fluxes by AMS-02 are actually the sum over the isotopes
(protons and deuterons,  and $^3$He and $^4$He). The model spectra used in the $\chi^2$
are thus also summed over the isotopes for consistency. }
In the above $\chi^2$ we make the simplifying assumption that all the data points are uncorrelated.
This is unlikely, since in most of the rigidity range the AMS-02 errors are dominated by
systematic uncertainties rather than statistical ones, and systematic uncertainties are
correlated in energy in various ways.
A more rigorous treatment would require a deeper knowledge of the various systematic
uncertainties and a way to model them, which, however,  requires a detector-level
analysis not available to us.
\ac{We, nonetheless, note that neglecting the correlations should imply larger errors in 
the CR parameters estimated from the fit, and it should thus correspond typically to a conservative assumption.}

The scan is intrinsically Bayesian, in the sense that \textsc{MultiNest} explores the posterior, which is
specified by the  likelihood $\mathcal{L}$ and the priors of the fitted parameters.
Nonetheless, if the posterior and its tails have been sampled accurately enough,
the likelihood samples collected by \textsc{MultiNest} can be also used for
a frequentist analysis. This typically requires more aggressive \textsc{MultiNest} settings
for a more accurate sampling of the likelihood.
We will use as default the frequentist interpretation of the scan in terms of the profile likelihood \cite{Rolke:2004mj},
providing a comparison with the Bayesian interpretation in a specific case.
As default, each scan is performed using for the  \textsc{MultiNest} settings
 $400$ live points, an enlargement factor  \texttt{efr}$=0.6$ and a tolerance  \texttt{tol}$=0.1$.
We verified that the results are stable varying these settings.
For each of the fits that   we will describe in the following the typical number of likelihood evaluations
performed by \textsc{MultiNest} is about 150,000.  At the same time, with the settings described in \secref{theory}
a \textsc{Galprop} run requires about 45 CPU-seconds, for a final total computational resources usage of about  $\sim$3 CPU-months.
The final efficiency (number of accepted steps over computed ones) of a typical scan is found to be $\sim$7\%. 
In the following, contour plots for two-dimensional profile likelihoods will be shown
at  the 1, 2, and 3 $\sigma$ confidence levels calculated from a two-dimensional $\chi^2$ distribution.   
The error in the single parameters will be calculated from 
 the related one-dimensional profile likelihoood and will be quoted at 1 $\sigma$ confidence level,
 i.e., from the condition $\Delta \chi^2=1$ with respect to the minimum $\chi^2$. 
 We will also show 1 and 2 $\sigma$ error bands around the best-fit spectra.
 They are derived from the envelope spectra of all the models lying within the  
 1 and 2 $\sigma$ best-fit region in the full multidimensional parameter space.

In total we perform fits  with up to 11 parameters, which can be grouped in two categories. 
The first one includes the  parameters of the shape of the injection spectrum: 
$\gamma_{1,p}$, $\gamma_{2,p}$, $\gamma_1$, $\gamma_2$,  $R_0$, and $s$. 
They denote the spectral indices, respectively,  for protons and for the heavier species below and above the break at $R_0$
 with smoothing $s$ (cf. \eqnref{SourceTerm_2}). 
As shown in the next section, the freedom in the individual spectral indices for protons, denoted with the subscript $p$, 
is necessary to achieve a good description of the measured data. 
We provide a dedicated study investigating the limits of a possibly universal injection spectrum at the beginning of the next \secref{results}. 
The second category includes the  parameters constraining the propagation, namely, 
the normalization $D_0$ and the slope $\delta$ of the diffusion coefficient, 
the Alfven velocity $v_\mathrm{A}$ related to reacceleration, 
the convection velocity $v_{0,\mathrm{c}}$, and the halo size $z_h$.
The above parameters are nonlinear and a new \textsc{Galprop} run has to be performed for every new parameter set.
On the other hand the fit includes three more parameters that  do not require a new \textsc{Galprop} run (for fixed values of the previous nonlinear parameters),
namely, $A_\mathrm{p}$, $A_\mathrm{He}$,  and $\phi_\mathrm{AMS}$, i.e., 
the normalization of the proton and helium fluxes as well as the solar modulation potential of AMS-02. 
For short we will call these parameters the \emph{linear parameters}, even though $\phi_\mathrm{AMS}$ does not act exactly linearly.
\ac{In principle, these parameters can be treated in the same way as the other, and this would give
a 14-dimensional parameter space to explore.
We can, however, exploit the fact that they do not require a \textsc{Galprop} evaluation to simplify the problem.
We, thus, do not include these parameters in the set of parameters scanned by \textsc{MultiNest},
but, instead,  we marginalize them on the fly for each set of the other 11  \textsc{MultiNest} parameters. 
More precisely, for each nonlinear parameter set sampled by \textsc{MultiNest} we 
search for the minimum $\chi^2$ over the linear parameters, and
we use this value to calculate the \textsc{MultiNest} likelihood.
In this way, we use \textsc{MultiNest} to effectively scan over the 11-dimensional
space which would be obtained from the 14-dimensional one marginalizing over the three linear parameters.}

Any deviation of the normalizations $A_i$ from $1$ implies a preference for a change in the CR species abundance
with respect to the input value.
As the normalization and spectrum of secondaries are calculated from the primaries'  input relative abundances,
rather than the ones rescaled by $A_i$, the input abundances
need to be adjusted if the best fit prefers  values of $A_i$ significantly different from 1.
We thus adjust the input abundances   iteratively for all fits including antiprotons,
repeating the fit until the normalizations $A_i$ converge to 1.
In practice, since the initial abundances are already very close to the 
ones preferred by the fit, only 1 or 2  iterations are typically required for convergence.
To this purpose, the output proton spectrum  from \textsc{Galprop} is normalized to a value
of  $4.4\cdot10^{-9}\,\mathrm{cm^{-2}s^{-1}sr^{-1}MeV^{-1}}$ 
    at  a kinetic energy of $100\,$GeV, and the parameter  $A_{\rm p}$ is thus relative to this value.
 The  helium ($^4$He) spectrum is instead normalized to  a final (i.e., found after the iterations) input abundance of 
 $7.80\cdot 10^4$ relative to a proton abundance of   $1.06\cdot 10^6$.
 The parameter $A_{\rm He}$ is thus relative to this normalization. 
 This value was found to be appropriate (i.e., giving $A_{\rm He}$ compatible with 1 after the fit)
 for all the fits performed, except for the case uni-PHePbar (see next section)
 where we used an input normalization of $9.48\cdot 10^4$.

\begin{table}[t!]
  \caption{List of \textsc{MultiNest} and linear parameters in the fit and respective ranges of variation. 
  \ac{See the text for a detailed description of the parameters.}}
  \label{tab::Parameters}
  \centering
  \begin{tabular}{l l r r l}
  \hline \hline
  \multicolumn{2}{l}{ \textbf{\textsc{MultiNest}} \textbf{parameters}} \hspace{0.5cm} & \multicolumn{3}{l}{\textbf{Ranges}}  \\ \hline \hline
  $\gamma_1           $  & $        $ & $1.2    $ & - & $2.3    $ \\ 
  $\gamma_2           $  & $        $ & $2.0    $ & - & $2.9    $ \\ 
  $\gamma_{1,p}       $  & $        $ & $1.2    $ & - & $2.3    $ \\ 
  $\gamma_{2,p}       $  & $        $ & $2.0    $ & - & $2.9    $ \\ 
  $R_0                $  & [GV]$    $ & $1.0    $ & - & $50     $ \\ 
  $s                  $  & $        $ & $0.05   $ & - & $1.0    $ \\
  $\delta             $  & $        $ & $0.1    $ & - & $0.9    $ \\
  $D_0                $  & [$10^{28}$ cm$^2$/s] & $0.5$ & - & $10.0$ \\ 
  $v_\mathrm{A}       $  & [km/s] $ $ & $0      $ & - & $60     $ \\ 
  $v_{0,\mathrm{c}}   $  & [km/s] $ $ & $0      $ & - & $100    $ \\ 
  $z_h                $  & [kpc]  $ $ & $2      $ & - & $7      $ \\ \hline \hline  
  \multicolumn{2}{l}{\textbf{Linear parameters}} & \multicolumn{3}{l}{\textbf{Ranges}}    \\ \hline \hline
  $A_\mathrm{p}       $ & $  $ &  $0.1$ &-& $5.0 $ \\
  $A_\mathrm{He}      $ & $  $ &  $0.1$ &-& $5.0 $ \\
  $\phi_\mathrm{AMS}  $ & [GV] &  $0  $ &-& $1.8 $ \\ \hline \hline
  \end{tabular}
\end{table}

Three further parameters, required to specify the CR model spectra, are kept fixed
or varied as a function of the other parameters. Specifically,
the smoothness transition parameter $s_1$ for the second break $R_1$ is kept fixed
to a value of 0.05, \ac{ given the sharp transition in this case, as can be seen directly in  the $p$ and He spectra.}
The break itself and index after the
break are fixed to $R_1= 450\,$GeV and $\gamma_3=\gamma_2-0.14$, for both $p$ and He.
The latter two parameters have been fixed with the following procedure: 
At large rigidities above $\sim 100\,$GV the CR spectrum is approximately given by the injection spectrum steepened by $\delta$,
\begin{eqnarray}
  \Phi_M(R) \sim q_R(R) \cdot R^{-\delta} \sim 
            \begin{cases}
               (R/R_1)^{-\gamma_2-\delta}\qquad R<R_1 \\
               (R/R_1)^{-\gamma_3-\delta}\qquad \text{else}
            \end{cases}.
\end{eqnarray}
Therefore it is possible to fit a broken power law directly  to the data in order to determine the break position $R_1$ 
and the amount of the break $\Delta \gamma = \gamma_3 - \gamma_2$. 
We performed two separate fits to $p$ only and He only data, using AMS-02 and CREAM data together. 
We find that both fits give compatible results for the break position and amount of break, with the values reported  above. 
The determination of  $R_1$  and $\Delta \gamma$ is where CREAM data play the main role in the analysis. 
For the general fits performed in the following, the weight of CREAM data is quite low, since the error
bars are very large compared to AMS-02. Indeed, we verified that excluding CREAM data from the fit
 did not significantly change the  fit constraints.
\ac{We also tested the case in which we exclude the CREAM data from the fit and for consistency we use
values of $R_1$  and $\Delta \gamma$  determined from AMS-02 only.  
In this case the same procedure described above gives $R_1=270$ GV  and $\Delta \gamma=-0.1$.
We find that the impact of these changes on the secondary antiproton spectrum is only at the level of 10-20\% of the 
error bars of the data points above 50 GV. The  fit constraints, consequently, are also not significantly affected.}

The fit parameters, linear and nonlinear, and their explored ranges are summarized in \tabref{Parameters}. 

\section{\label{sec::results}Results}

The results are presented as follows. We first discuss in \subsecref{UniversalInjectionSpectrum} the possibility of fitting the data with a universal injection spectrum.
In \subsecref{main}  we then proceed to the main analysis were separate spectral indices for the proton injection spectrum are allowed. 
In \subsecref{antiproton}  we discuss the impact of the uncertainties related to the antiproton production cross section.
A comparison between the frequentist and Bayesian results is shown in \subsecref{Bayes}. 
Finally, in \subsecref{1GV} and \subsecref{convection} we check the robustness of the analysis results with respect to  the chosen rigidity fit range and to the inclusion or exclusion of convection in the fit.

\subsection{\label{subsec::UniversalInjectionSpectrum}Universal injection spectrum}

\begin{figure}[b!]
	\centering
	\begin{subfigure}[b]{0.45\textwidth}
	  \includegraphics[width=\textwidth]{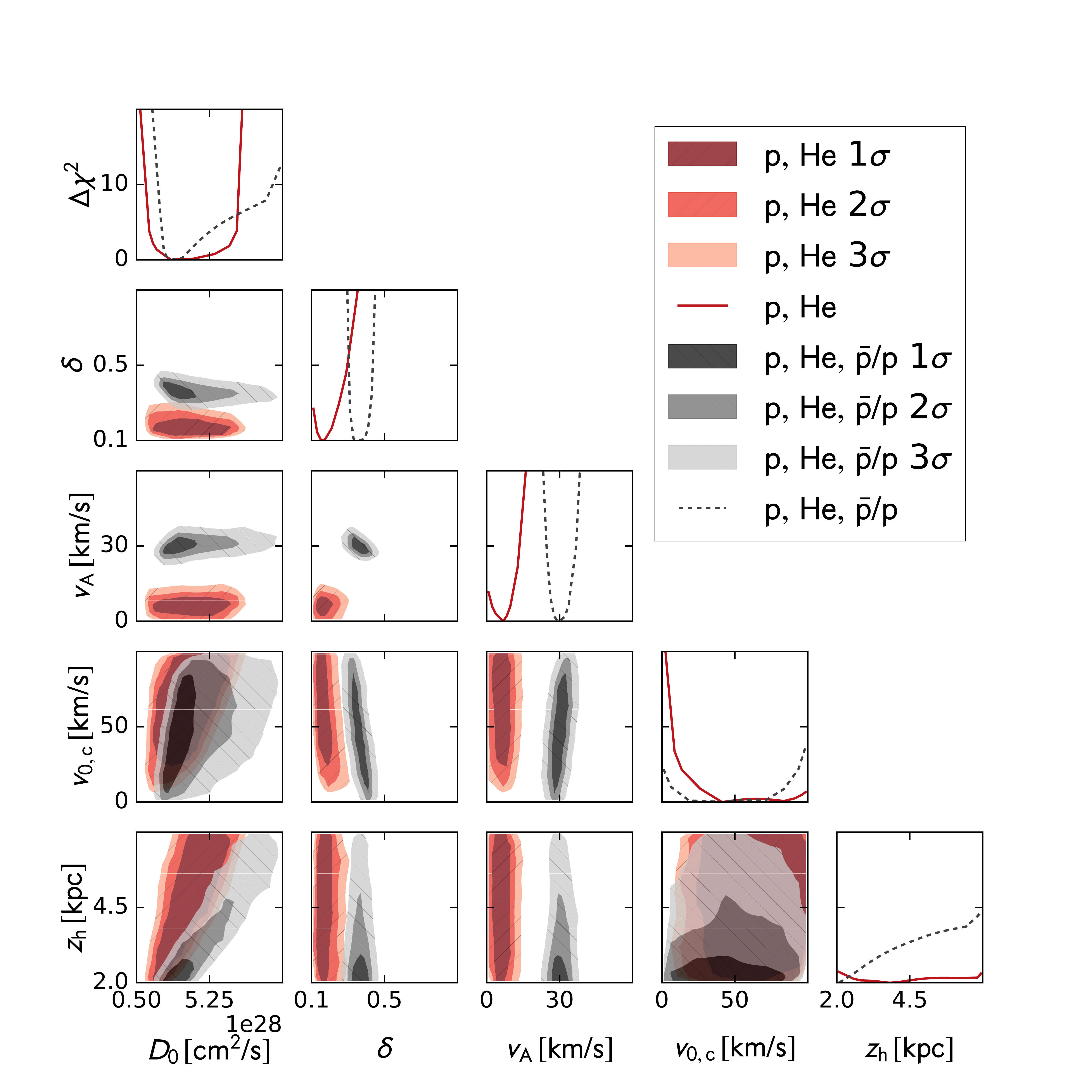}
	  \caption{Triangle plot for a selected set of propagation parameters.}
	  \label{fig::universalInjec_triangle}
	\end{subfigure}
	\hspace{0.02\textwidth}
	\begin{subfigure}[b]{0.45\textwidth}
	  \includegraphics[width=\textwidth]{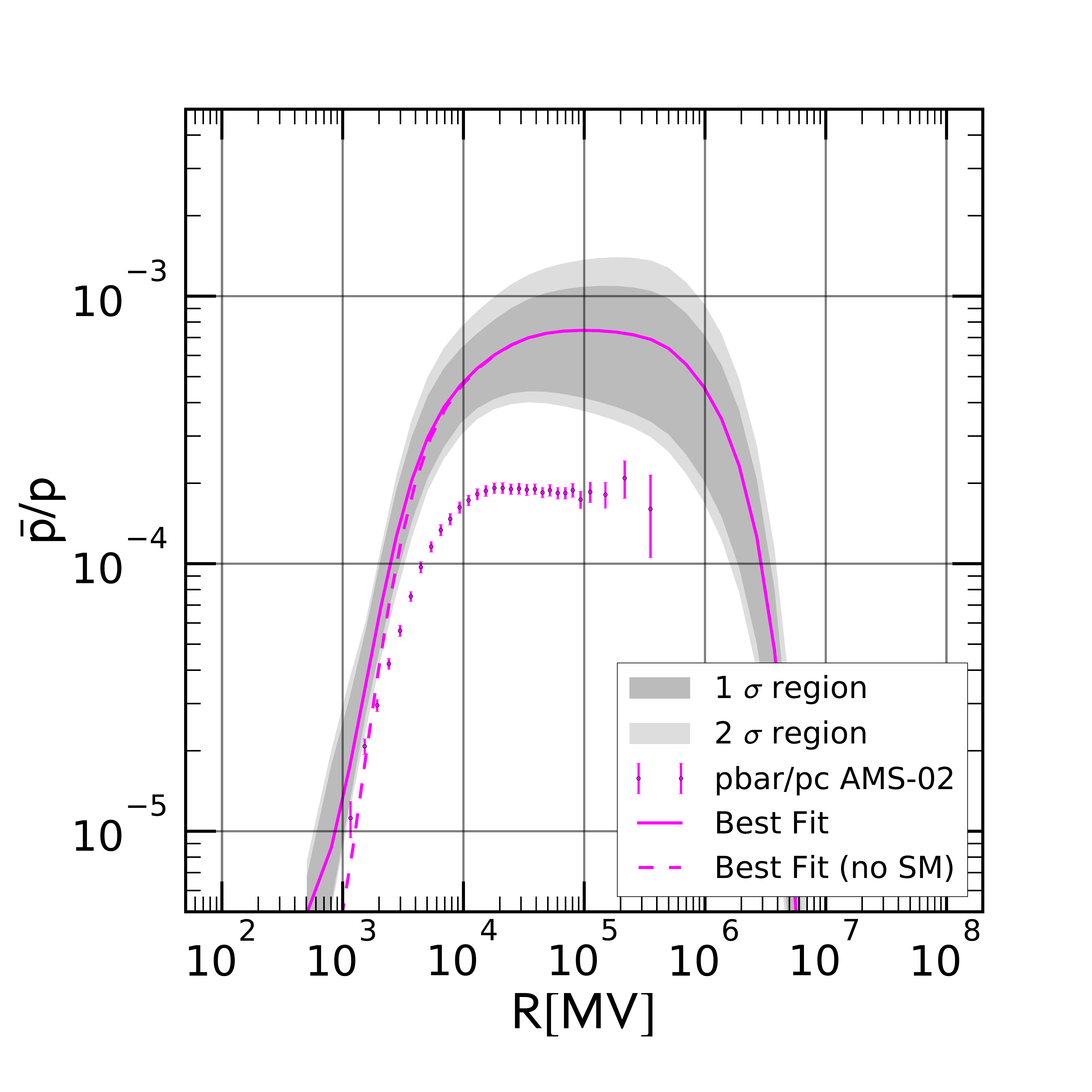}
  	\caption{Predicted antiproton spectrum for the (uni-PHe) fit. \\ \hspace{0.4cm}}
	  \label{fig::universalInjec_Pbar_prediction}
	\end{subfigure}
	\begin{subfigure}[b]{0.45\textwidth}
	  \includegraphics[width=\textwidth]{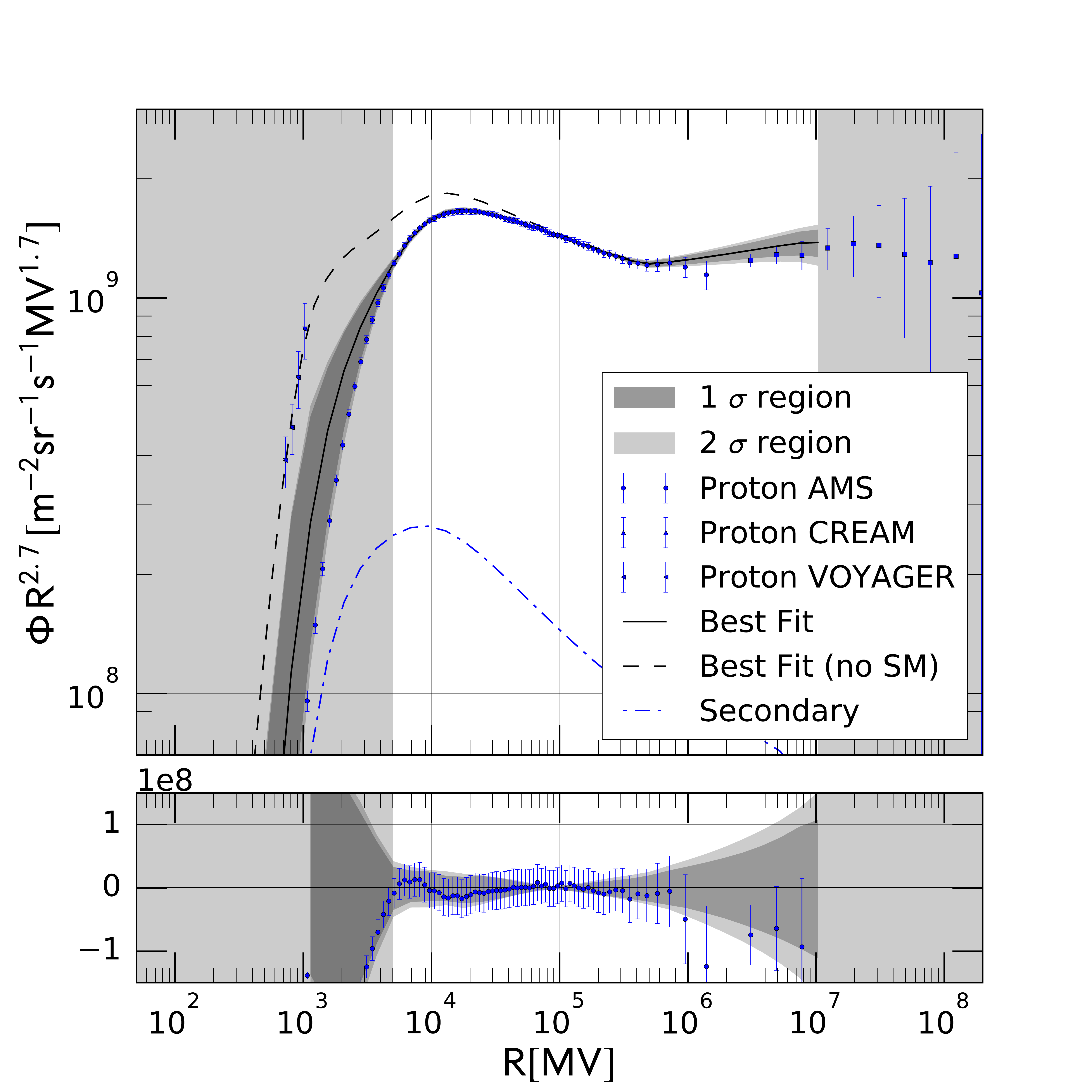}
  	\caption{Best fit results for protons for (uni-PHe).}
	  \label{fig::universalInjec_P_PHe}
	\end{subfigure}
	\hspace{0.02\textwidth}
	\begin{subfigure}[b]{0.45\textwidth}
    \includegraphics[width=\textwidth]{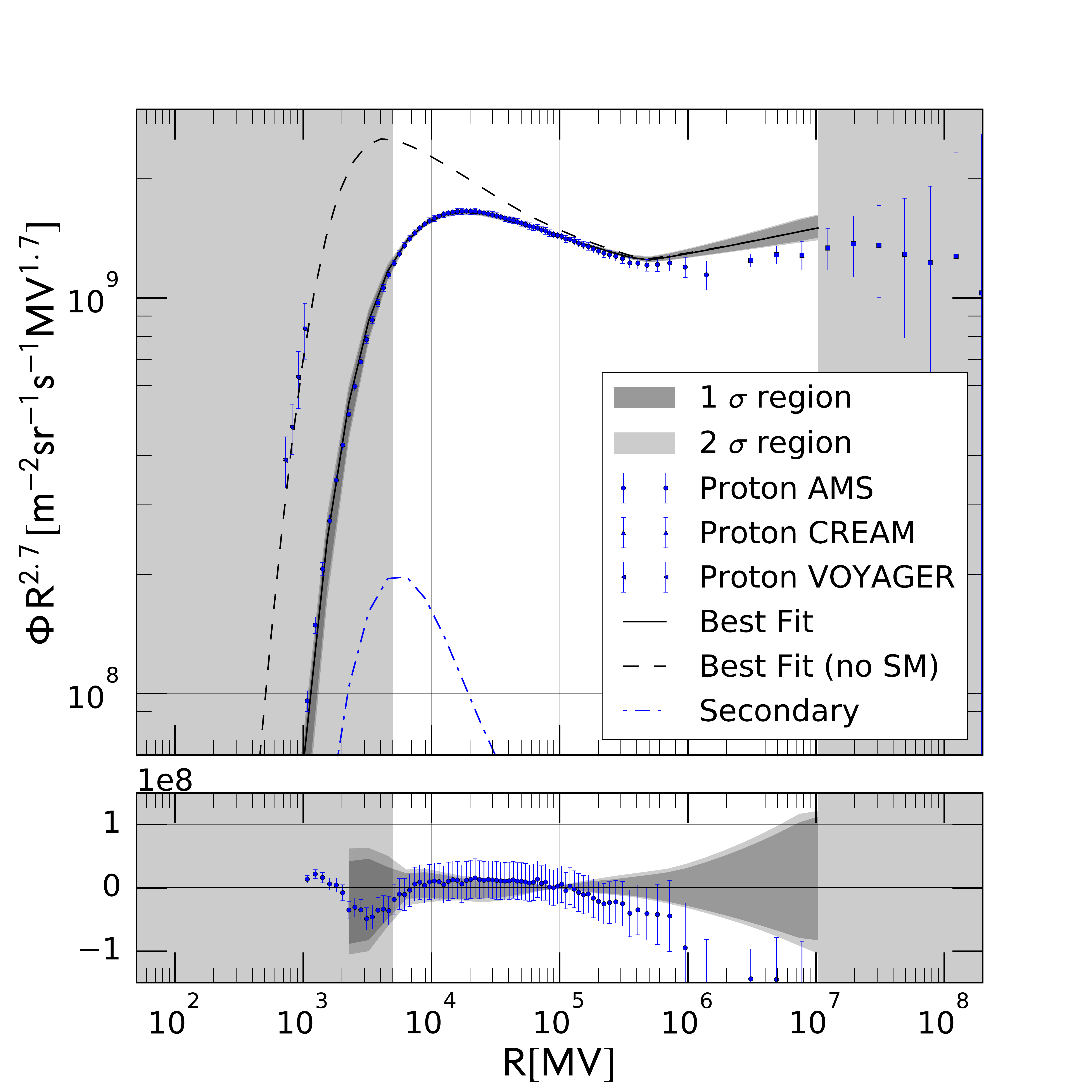}
  	\caption{Best fit results for protons for (uni-PHePbar).}
	  \label{fig::universalInjec_P_PHePbar}
	\end{subfigure}
	
	\caption{Comparison of the fit to data set (uni-PHe) and (uni-PHePbar) in the framework of a universal injection spectrum.
                In (c) and (d) the light-gray shaded regions indicate the rigidity range used in the fit.
	              The lower panels in each of the plots show the residuals with respect to the best fit.}
  \label{fig::universalInjec}
\end{figure}

One of the unexpected features revealed by the precise measurements of 
PAMELA and AMS-02  is a significant discrepancy of the proton and helium spectral indices above $\sim 30\,$GV,
with $\Delta\gamma_{p,\mathrm{He}} = 0.101 \pm 0.014$ (PAMELA,  \cite{Adriani_PAMELA_pHe_2011}) and
$\Delta\gamma_{p,He} = 0.077 \pm 0.007$ (AMS-02,  \cite{Aguilar_AMS_Proton_2015,Aguilar_AMS_Helium_2015}).
From the theoretical point of view the reason for this difference is unclear and various possibilities have been discussed
\cite{Vladimirov:2011rn,Serpico:2015caa,Blasi:2011fi,Malkov:2011gb,Ohira:2015ega,Ohira:2010eq}.
Moreover, acceleration in the sources above $\sim 30\,$GV is expected to be 
charge independent and therefore the same universal injection index is expected for $p$ and He,
as well as for the other species.
We thus first investigate the possibility to fit the data assuming a universal injection index, 
attributing the difference in the observed indices to propagation effects.
More in detail, we perform the fit as described in  \secref{methods}, 
but we force the injection spectrum of protons and helium to be equal, i.e. $\gamma_{1,p}=\gamma_1$ and $\gamma_{2,p}=\gamma_2$,
thus reducing the parameter space from 11 to 9 dimensions.

The fit is performed with two different data sets: in one case using only protons and helium (fit labeled as uni-PHe),
and in the second case using proton, helium, and the antiproton-to-proton ratio  (fit labeled uni-PHePbar). 
Results are shown in \figref{universalInjec}. 
In the (uni-PHe) case we
obtain a good fit with a minimal $\chi^2/$ of 53.1 for a number of degrees of freedom (NDF) of 124. It can be seen in \figref{universalInjec_P_PHe} that the best fit residuals with respect to the proton data are very flat in the fitted rigidity range.
A similar result is obtained for the helium spectrum (not shown).
The difference in the index between $p$ and He is explained by a significant production of secondary protons
that soften the observed total (primaries plus secondaries) proton spectrum   by the required $\sim$0.1 value,
with respect to the helium spectrum.
In turn this imposes strong constraints in the diffusion parameter space, as can be seen by the red
contours in \figref{universalInjec_triangle}. In particular, a low value of $\delta \sim 0.15$ and a low
amount of reaccelaration $v_\mathrm{A}\sim 0$  are required. 
Although this scenario is appealing, it is ultimately  revealed to be problematic.
A first problem is the amount of solar modulation required by the fit, given by  $\phi_{\rm AMS}= 300^{+60}_{-75}\,$MV,
which is quite low with respect to the neutron monitor 
expectation\footnote{An updated table for the solar modulation potential up to 2016 is available under \url{http://cosmicrays.oulu.fi/phi/Phi_mon.txt}.} 
of  $\sim$ 500-600$\,$MV \cite{Usoskin_Solar_Modulation_2011}.
The second, more severe, problem is the fact that antiprotons are significantly overpredicted with respect to the observations,
\ac{as shown in the lower panel of \figref{universalInjec_Pbar_prediction}}.
This can also be seen from the result of the (uni-PHePbar) fit.
In fact, the parameter space constraints from this fit,  shown by the black-gray contours in \figref{universalInjec_triangle},
select a much larger value of $\delta\sim 0.4$ and of $v_\mathrm{A}\sim 30\,$km/s,
which are incompatible, at high significance with the (uni-PHe) results.
With the higher $\delta$ antiprotons data are now correctly produced but the amount of secondary 
protons is not enough anymore to explain the proton-helium index difference. 
This can clearly be seen from the systematic behavior of the residuals in \figref{universalInjec_P_PHePbar},
despite the fact that, formally, the fit is still reasonable, with a $\chi^2/$NDF of $140.4/147$.
The (uni-PHePbar) fit also provides a more reasonable amount of solar modulation, with $\phi_{\rm AMS}= 780^{+80}_{-40}\,$MV.
These results are in qualitative agreement  with 
\cite{Vladimirov:2011rn,Blasi:2011fi}, where He spallation effects were studied to explain the difference
in slope between proton and He, and similar difficulties in explaining secondaries spectra were encountered.
\ac{Note also that, as explained in \secref{theory}, we are neglecting the eventual contribution of secondary protons
from $Z>$2 nuclei. This is not expected to be crucial, since, even in the case this contribution would be large (20-30\%) 
this would not solve the above issues.  }

In the light of the above results, we will adopt in the following as the main scenario the one in which
the proton and helium spectral indices are varied independently.

\subsection{\label{subsec::main} Main fit}


Using the 11-dimensional setup discussed in \secref{methods}, we will now perform fits
to different data sets, to test the self-consistency of the results.
In particular, we consider the following 3 fits: using only proton data (P),  
proton and helium data  (PHe), and  proton, helium and antiproton data (main).
\figref{dataSetComparison_triangle} shows how the propagation parameter space  successively shrinks by going from data set (P) to (main). 
As expected, because of the large degeneracy of the parameters in case (P) nearly the whole sampled parameter space is allowed. 
Adding helium data results in a tendency against reacceleration, a  preference towards large values for the convection velocity  $v_{c,0}\agt 50\,$km/s,
and a diffusive halo height $z_h\agt 4\,$kpc.
The constraints, however, are not extremely strong, and at the $\sim 3 \sigma$ level again almost the whole parameter space
is allowed. 
\begin{figure}[b!]
	\centering
	\begin{subfigure}[b]{0.48\textwidth}
	  \includegraphics[width=\textwidth]
	    {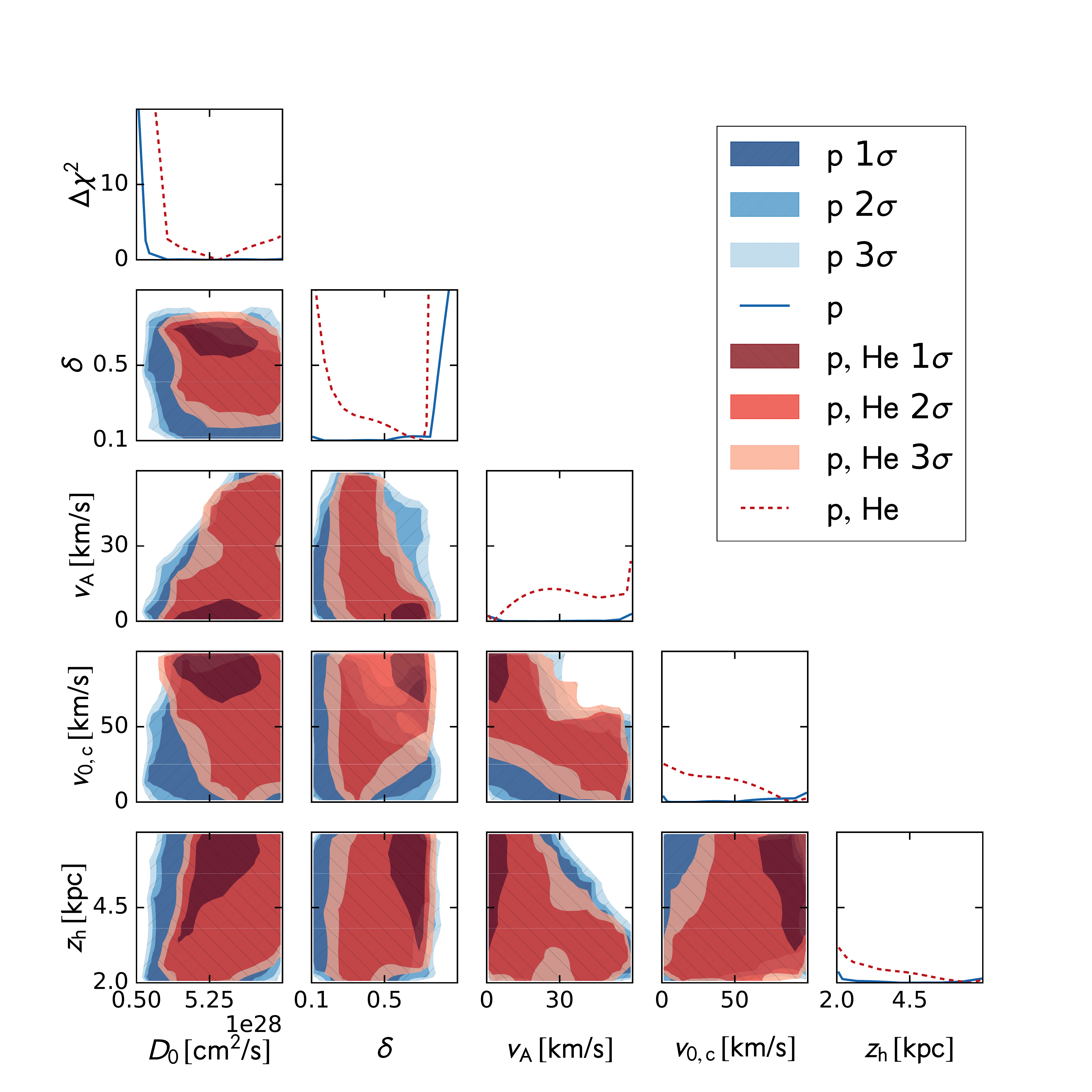}
  	\caption{(P) vs. (PHe)}
	  \label{fig::5GV_P_comp_PHe_triangle}
	\end{subfigure}
	\begin{subfigure}[b]{0.48\textwidth}
	  \includegraphics[width=\textwidth]
	    {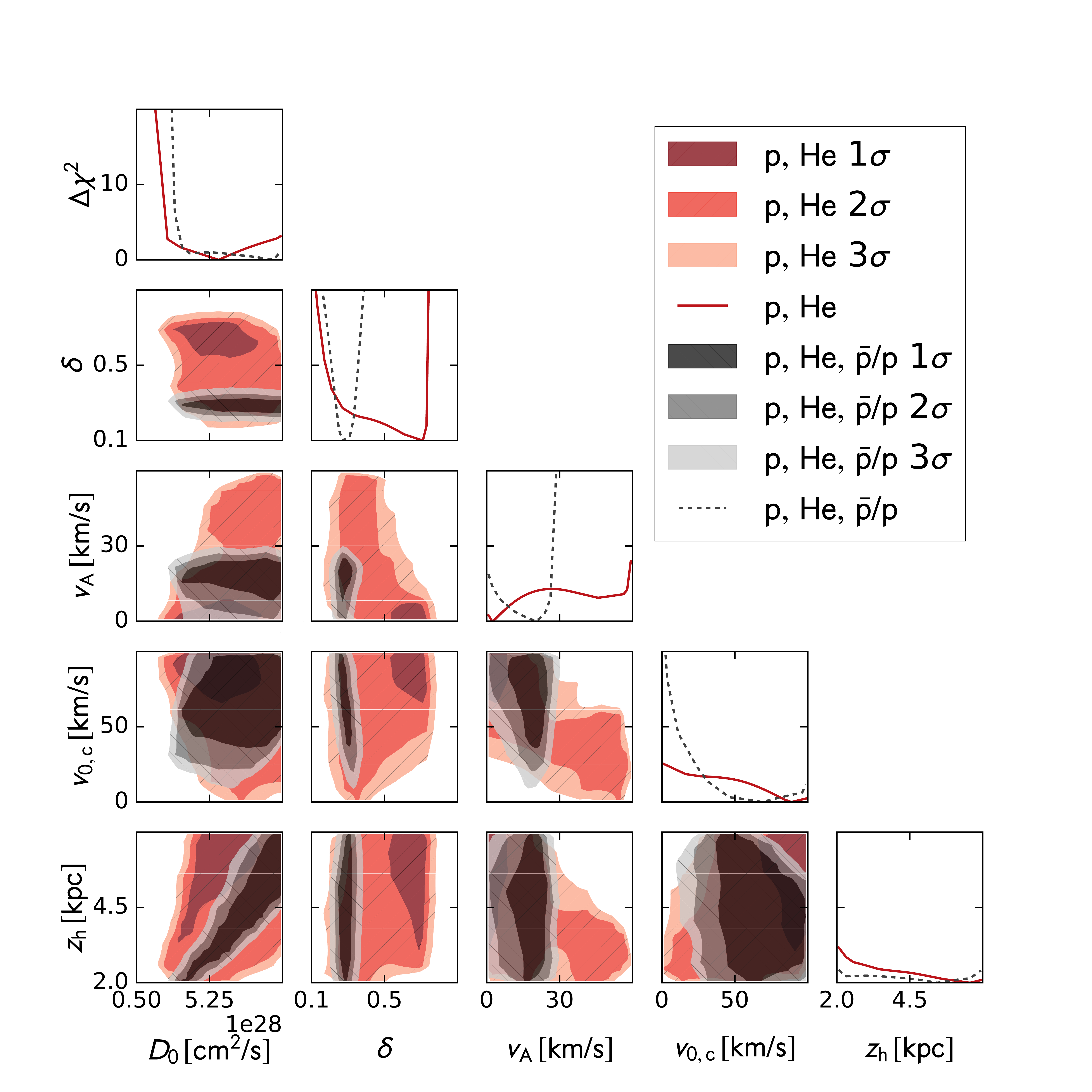}
    \caption{(PHe) vs. (main)}
	  \label{fig::5GV_PHe_comp_PHePbar_triangle}
	\end{subfigure}
	\caption{Comparison of fit results for the three data sets (P), (PHe), and (main) in the main 
	          fit framework (11 parameters) for a selected set of propagation parameters. }
  \label{fig::dataSetComparison_triangle}
\end{figure}

\begin{figure}[b!]
  \begin{minipage}{1.\textwidth}
  	\centering
  	\includegraphics[width=0.97\linewidth]{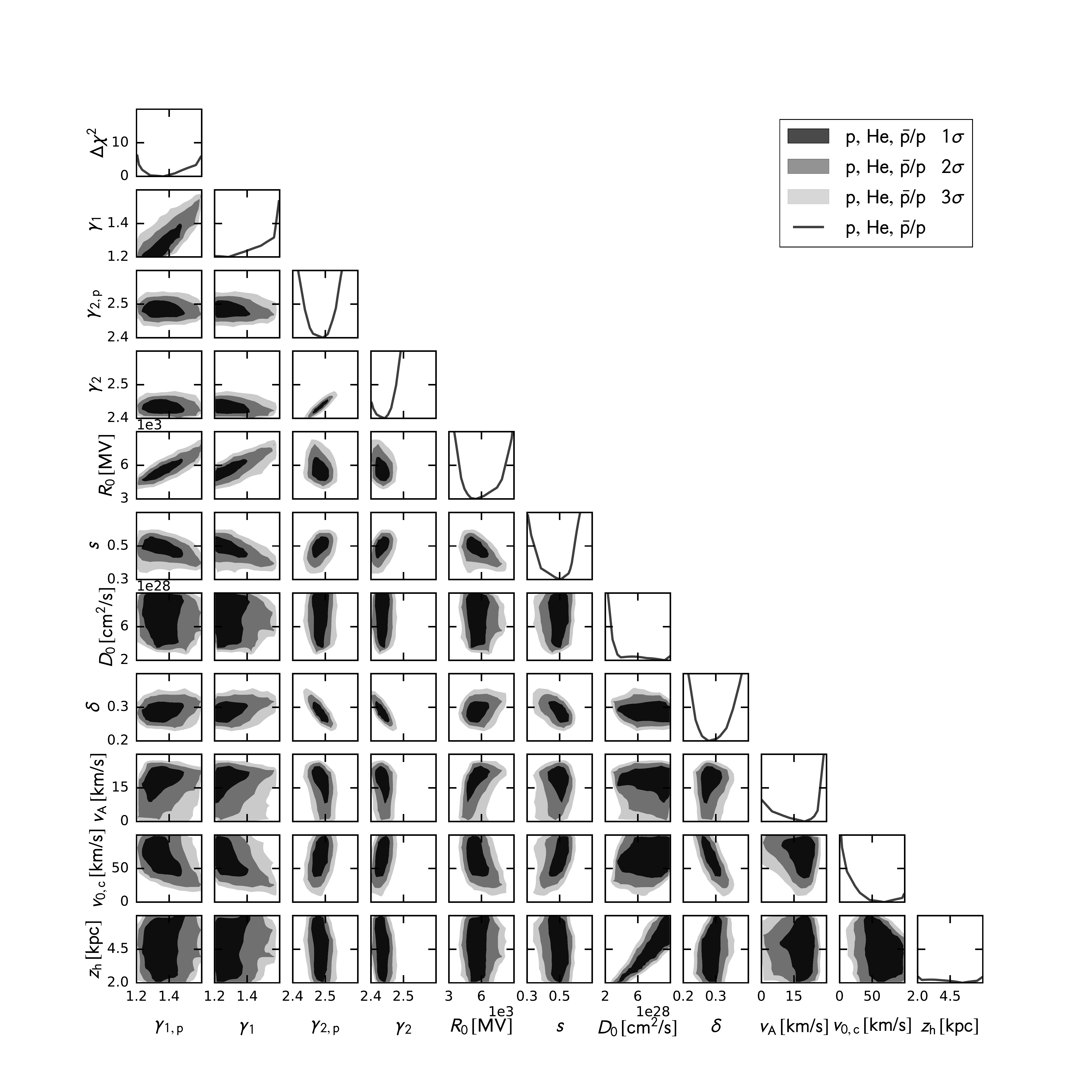}
  	\caption{Full triangle plot for the results of the main fit using  protons, helium and antiprotons (main).}
  	\label{fig::5GV_PHePbar_triangle}
  \end{minipage}	
	\begin{minipage}{.7\textwidth}
	  \centering
	  \includegraphics[width=0.25\textwidth]
	    {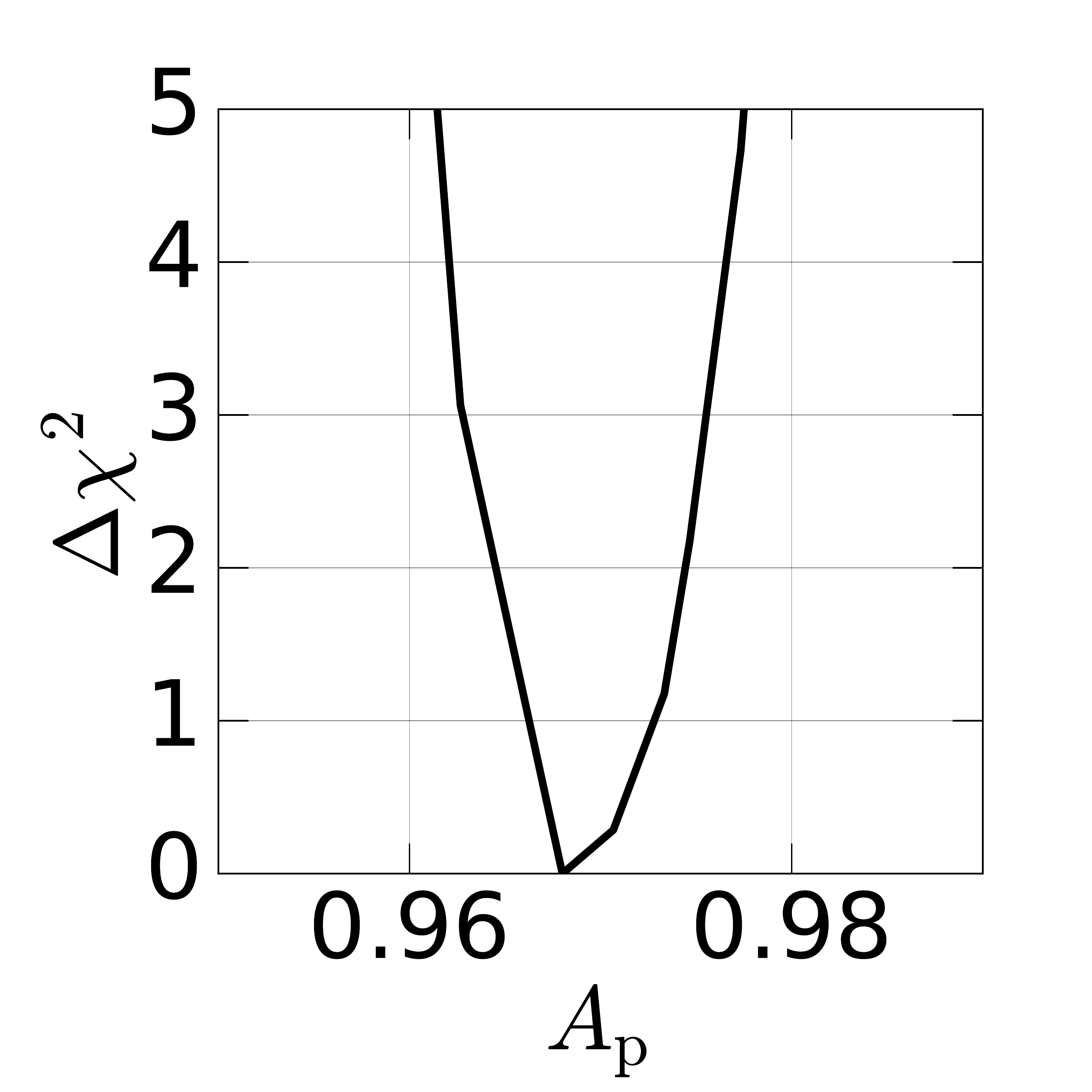}
	  \includegraphics[width=0.25\textwidth]
	    {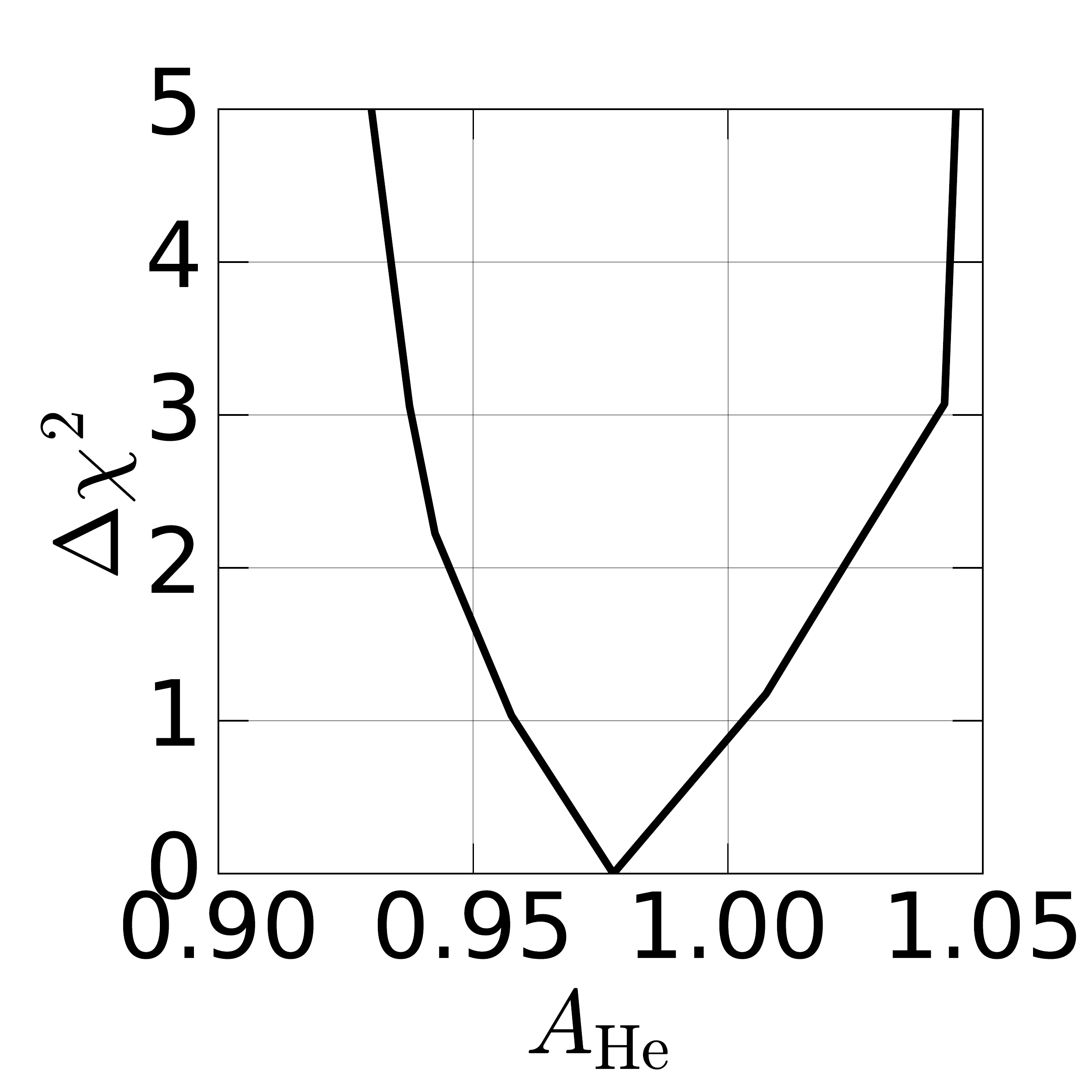}
	  \includegraphics[width=0.25\textwidth]
	    {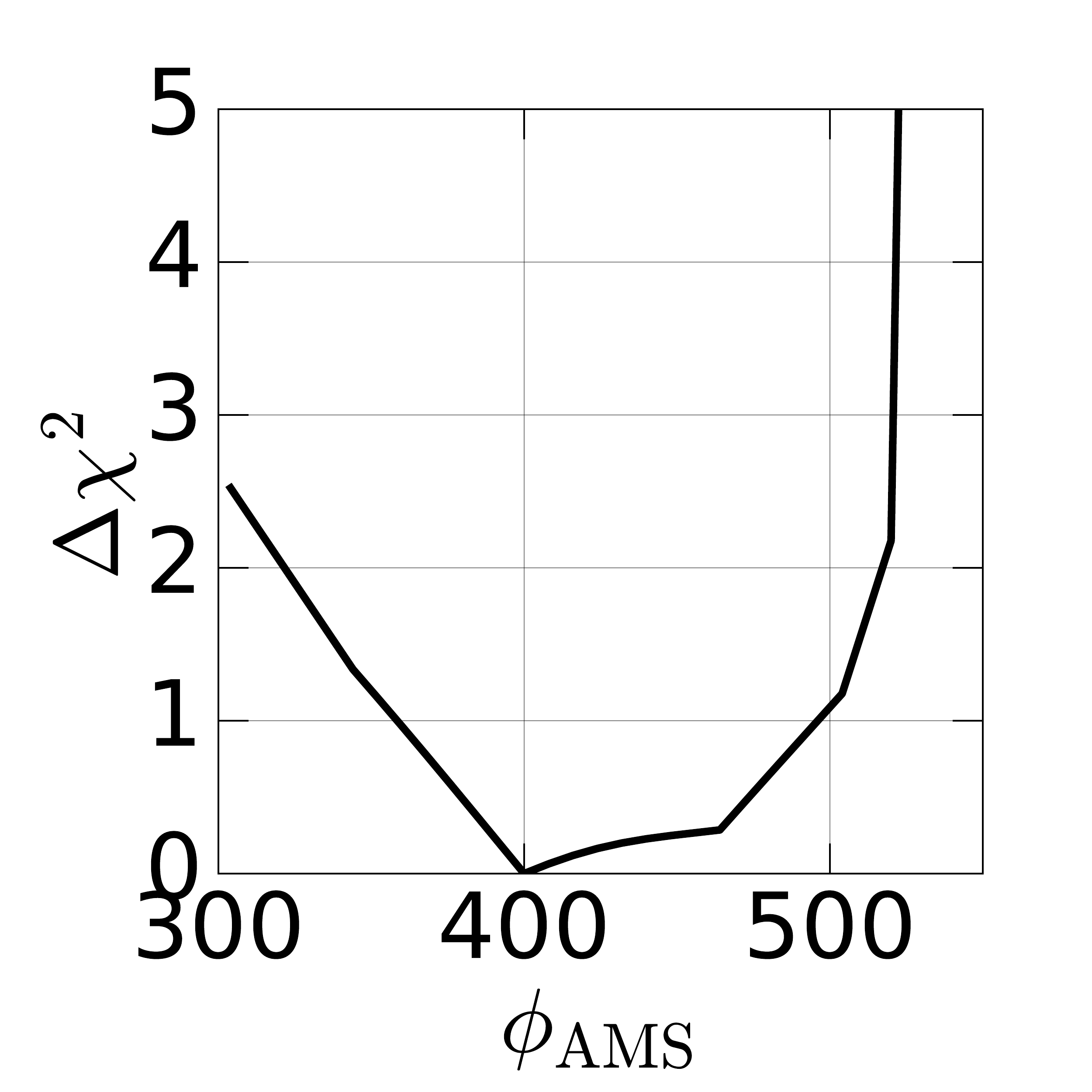}
	  \caption{Profiles of the linear parameters for the (main) fit.\\ \vspace{1\baselineskip}}
	  \label{fig::5GV_PHePbar_linearParameters}
	\end{minipage}
\end{figure}

\figref{5GV_PHe_comp_PHePbar_triangle} shows the comparison between (PHe) and (main) results.
As expected the secondary antiprotons give tight constraints on  the rigidity dependence of diffusion $\delta \sim 0.3$,
while the usual degeneracy in $D_0$-$z_h$ appears, and no constraints on $z_h$ can be inferred. This is also expected since  strong constraints on $z_h$  can be achieved only using precise data on radioactive clocks like $^{10}$Be/$^{9}$Be,
which are not yet available.
Finally, $v_{\rm A}$ and $v_{0,c}$ are poorly constrained individually 
apart from a tendency to not prefer strong reacceleration  $v_\mathrm{A}\alt 30\,$km/s
and a favor for large convection  $v_{0,c}\agt 50\,$km/s. 
This is mainly due to the fact that they have approximately degenerate effects on the spectra, and only 
a combination of the two parameter is well constrained.
Interestingly, a fit with only convection and no reacceleration seems, thus, possible. 
Indeed, some critical view on reacceleration has been recently discussed \cite{Drury:2015zma}.
It can be seen that (PHe) and (main) are not compatible at the $2\sigma$ level,
although they became fully compatible at the $3\sigma$ level.
Given the very small error bars of AMS-02 it is perhaps expected that incompatibilities at the
$2\sigma$ level might appear,  due to the fact that the level of complexity
of the fitted models is likely starting to be not comparably adequate.
We thus deem the compatibility at $3\sigma$ reasonable.

\begin{figure}[t!]
	\centering
	\begin{subfigure}[b]{0.4\textwidth}
	  \includegraphics[width=\textwidth]{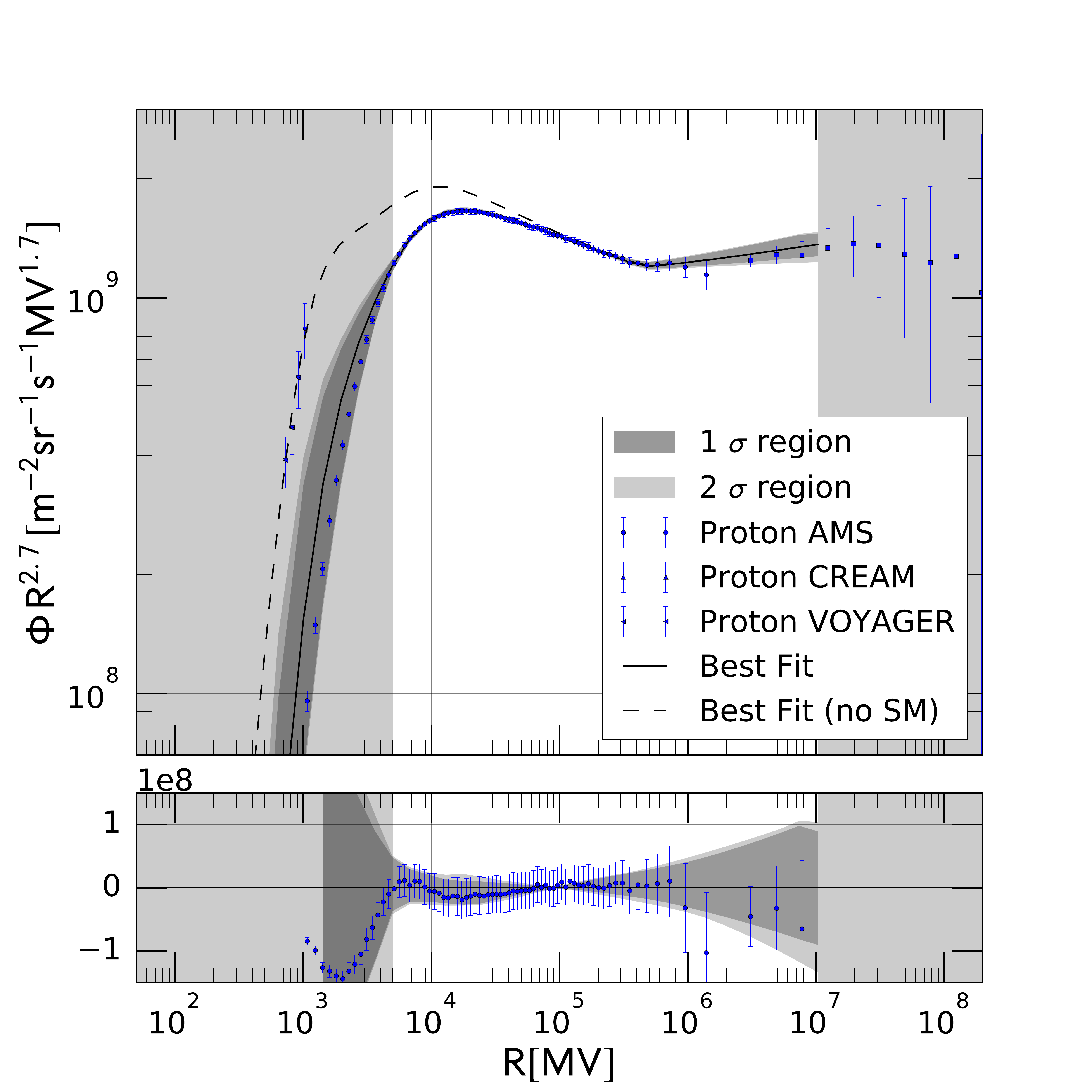}
  	\caption{Proton}
	  \label{fig::PHePbar_bestFit_P}
	\end{subfigure}
	\begin{subfigure}[b]{0.4\textwidth}
	  \includegraphics[width=\textwidth]{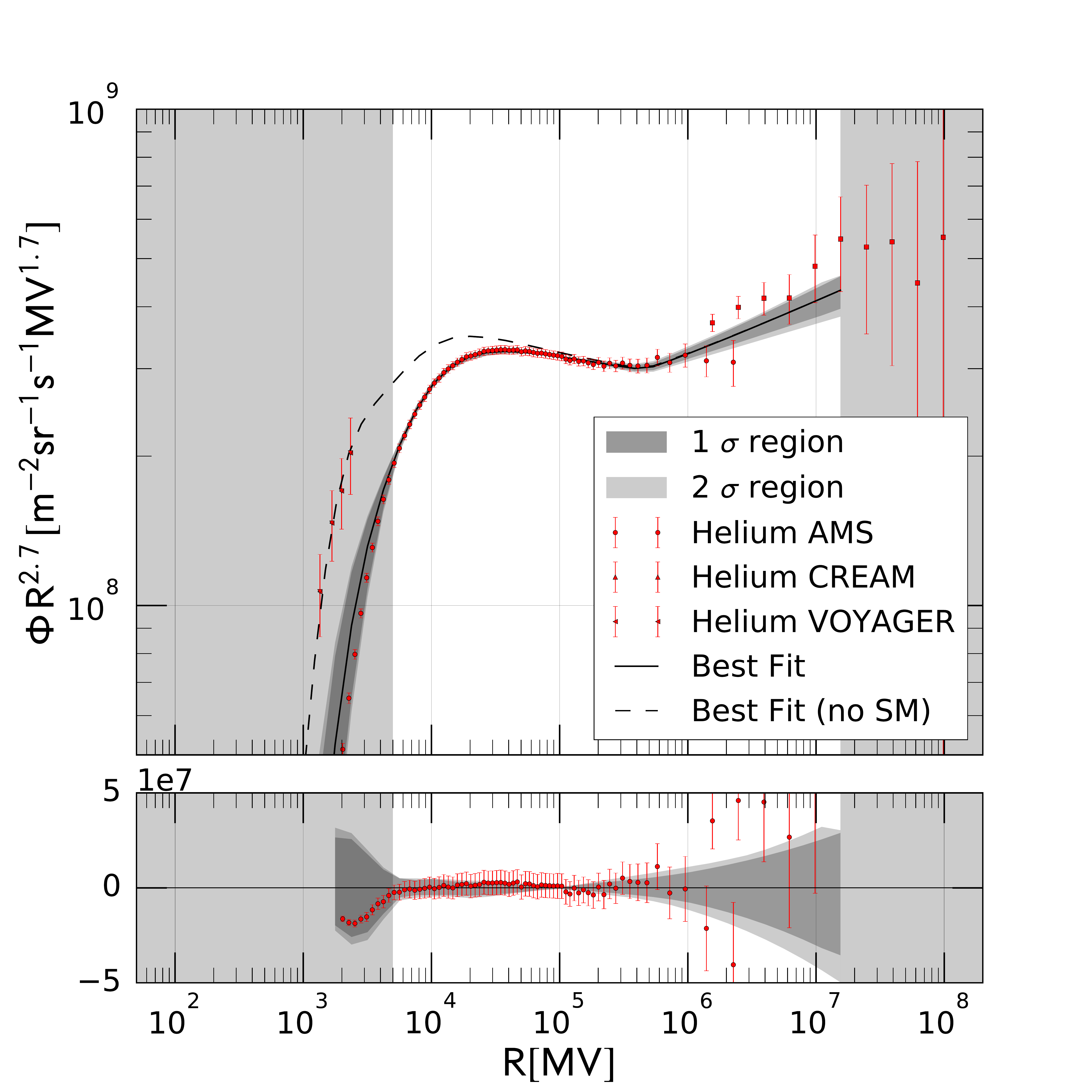}
    \caption{Helium}
	  \label{fig::PHePbar_bestFit_He}
	\end{subfigure}
	\caption{Comparison between data and best-fit model for the main fit framework (main): 
	            11 parameters and fit to 
              proton, helium, and antiprotons. }
	\label{fig::PHePbar_bestFit}
	\vspace{1cm}
\end{figure}

\begin{figure}[t!]
	\begin{minipage}{.57\textwidth}
	\centering
	\includegraphics[width=1.0\textwidth]
    {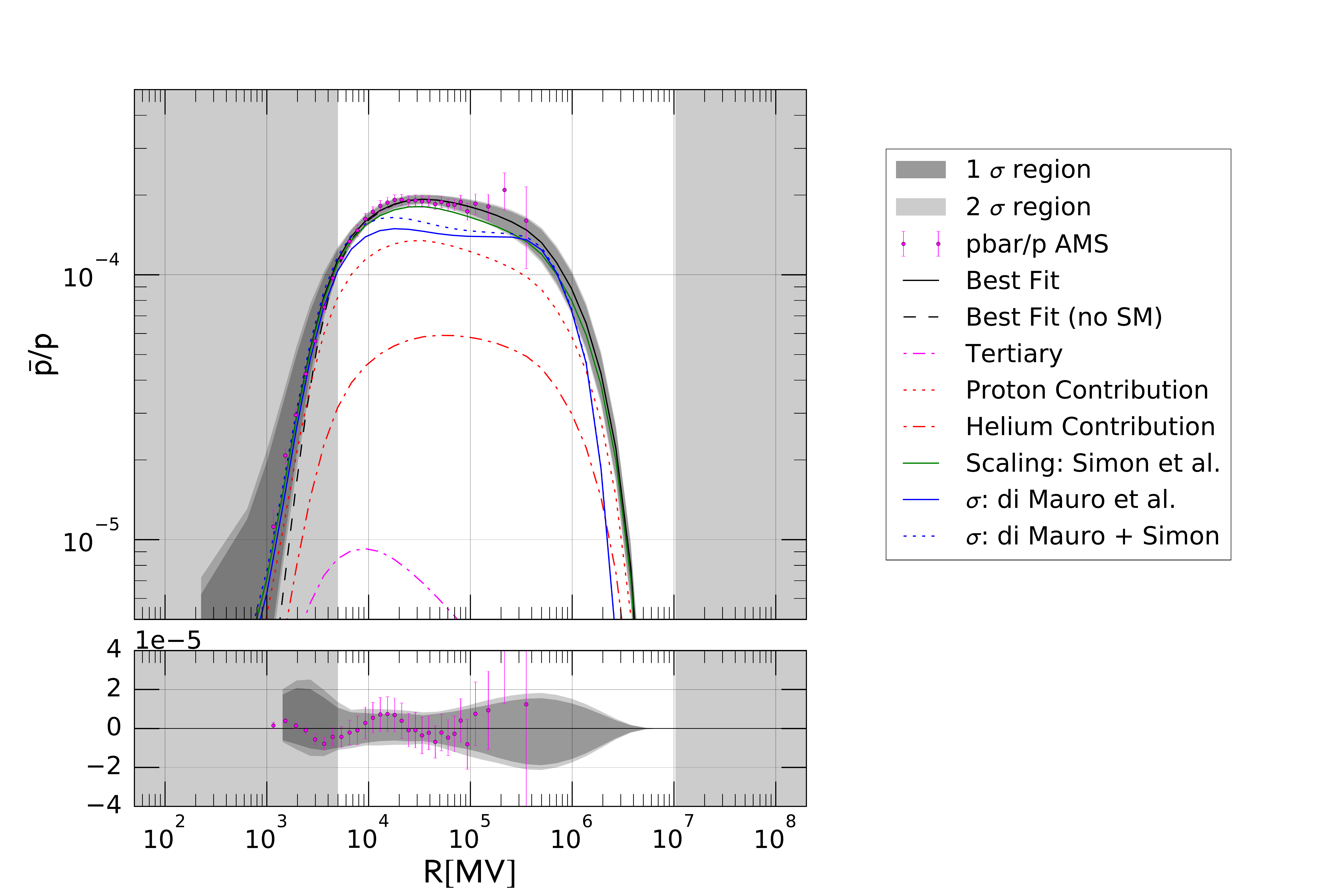}
  \caption{Comparison between ${\bar{p}/p}$ data and best-fit model for the main fit (main) framework 
              (11 parameters and fit to 
              proton, helium, and antiprotons). The various contributions to the total antiproton spectrum are also shown,
              as well as different production cross section models.
              The proton contribution includes only $pp$ production, whereas the helium contribution also 
              includes $p\,$He and He$\,p$ production.}
  \label{fig::5GV_PHePbar_pbar}
  \end{minipage}
	\hspace{.03\textwidth}
  \begin{minipage}{.38\textwidth}
  \includegraphics[width=1.\textwidth]
	    {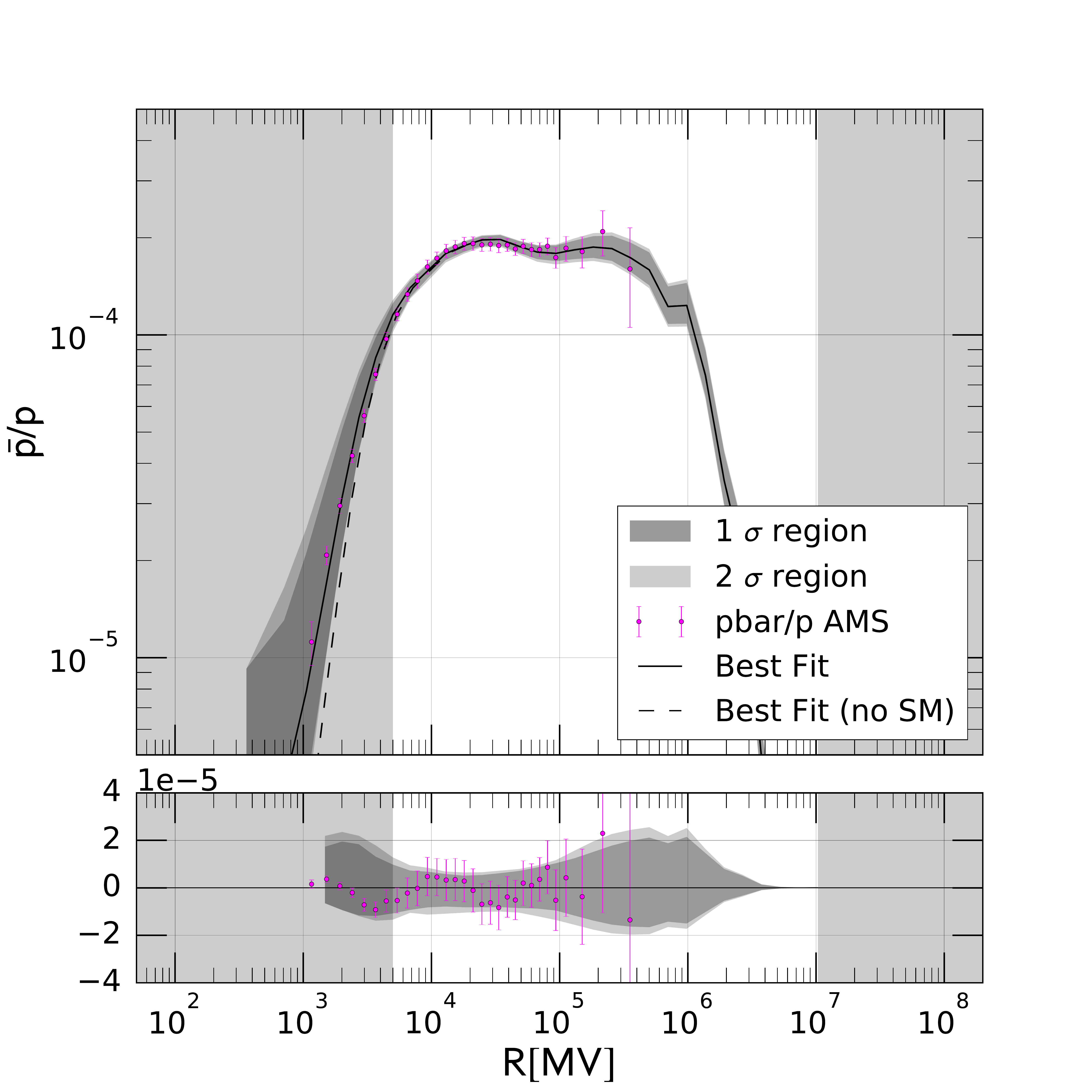}
  \caption{$\bar{p}/p$ ratio for fit using the antiproton production cross section from 
                  di Mauro et al. \cite{Mauro_Antiproton_Cross_Section_2014}.\\ \vspace{5.2em} }
	\label{fig::5GV_PHePbar_diMauro_CS}
	\end{minipage}
\end{figure}

For the case (main) the complete parameter space is shown in \figref{5GV_PHePbar_triangle} and 
the likelihood profiles of the linear parameters are given in \figref{5GV_PHePbar_linearParameters}. 
It can be seen that indeed $A_{\rm He}$ is compatible with 1 at 1$\sigma$. $A_{\rm p}$ 
does actually differ from 1, but only at the 3\% level which is much smaller than the uncertainties
in the other parameters. We thus did not perform a further fit iteration, readjusting again the input proton normalization.
The minimal $\chi^2/$NDF of the best fit point is $39.0/145$. 
The agreement between data and model is thus very good (cf. also \figref{PHePbar_bestFit} and \figref{5GV_PHePbar_pbar}). 
There are no systematics features  in the residuals of proton and helium spectra.  
The small residual structures in the antiproton-to-proton ratio are within the error band. 
The best-fit parameter values and their uncertainties are summarized in \tabref{BestFit}.
Finally, also the VOYAGER $p$ and He measurements are well fitted
by the unmodulated model spectra, as shown in \figref{PHePbar_bestFit}.
The best fit for the position of the break $R_0$ is compatible with $5\,$GV, our lower rigidity threshold,
indicating that a low rigidity break is not necessary to fit the data. We will comment more on
this point in \subsecref{1GV} where we show the results of the fit including data down to $1\,$GV.

The effect on the fit of the parameter $s$ introduced in this work can be inferred from \figref{5GV_PHePbar_triangle}. 
It can be seen that, apart from the expected degeneracy with the break position, $s$ has only mild degeneracies
with the other parameters.
Indeed, performing explicitly a fit without $s$ (a sharp break),
we found that only the ranges for $v_{\rm A}$ and $v_{0,c}$ slightly change, the two
parameters being in any case not well determined.
The main effect of $s$ is, instead, to provide an overall  better fit to the data and flatter residuals.
It is unclear if the need for $s$ in the fit implies, indeed,  that the injection spectra have a smooth break, or, alternatively
if  $s$ is compensating for a different effect, as, e.g.,  systematics in the modeling of the solar modulation.

In the following sections we will take (main) as baseline for further cross-checks and systematic studies.


\subsection{Antiproton production cross section}
\label{subsec::antiproton}

The lack of precise measurements of the antiproton production cross section (cf. \eqnref{SourceTerm_pbar}) 
constitutes an important systematic uncertainty in the interpretation of the precisely measured fluxes 
\cite{Donato:2001ms,Mauro_Antiproton_Cross_Section_2014}. 
Detailed measurements for the antiproton production exist only for proton-proton
inelastic scattering up to center of mass energies of $\sim$63 
GeV\footnote{Some sparse measurement up to $\sim$200 GeV also exists \cite{Mauro_Antiproton_Cross_Section_2014}.} 
\cite{Mauro_Antiproton_Cross_Section_2014}. 
For larger energies, or different target particles it is necessary to extrapolate and/or rescale the cross sections, 
leading to model dependent results.
Furthermore,  in the proton-proton inelastic scattering, only the antiproton production
cross section is directly measured, while no measurement is available for the antineutron 
(which subsequently decay into antiproton) production cross section.
In principle, from isospin symmetry the latter is expected to be equal to the former.
On the other hand a measurement from NA49 \cite{Fischer:2003xh} suggests that the antineutron
cross section is actually slightly larger than the antiproton one. 
Further details are discussed in refs. \cite{Mauro_Antiproton_Cross_Section_2014,Kappl:2014hha}.
In this section we compare our default choice of $p + p \rightarrow \bar{p} + X$ cross section given by Tan \& Ng \cite{TanNg_AntiprotonParametrization_1983} as implemented in
\textsc{Galprop} with the more recent study in \cite{Mauro_Antiproton_Cross_Section_2014}.

\begin{figure}[t!]
  \begin{minipage}{.45\textwidth}
	\centering
	\includegraphics[width=1.\textwidth]
	    {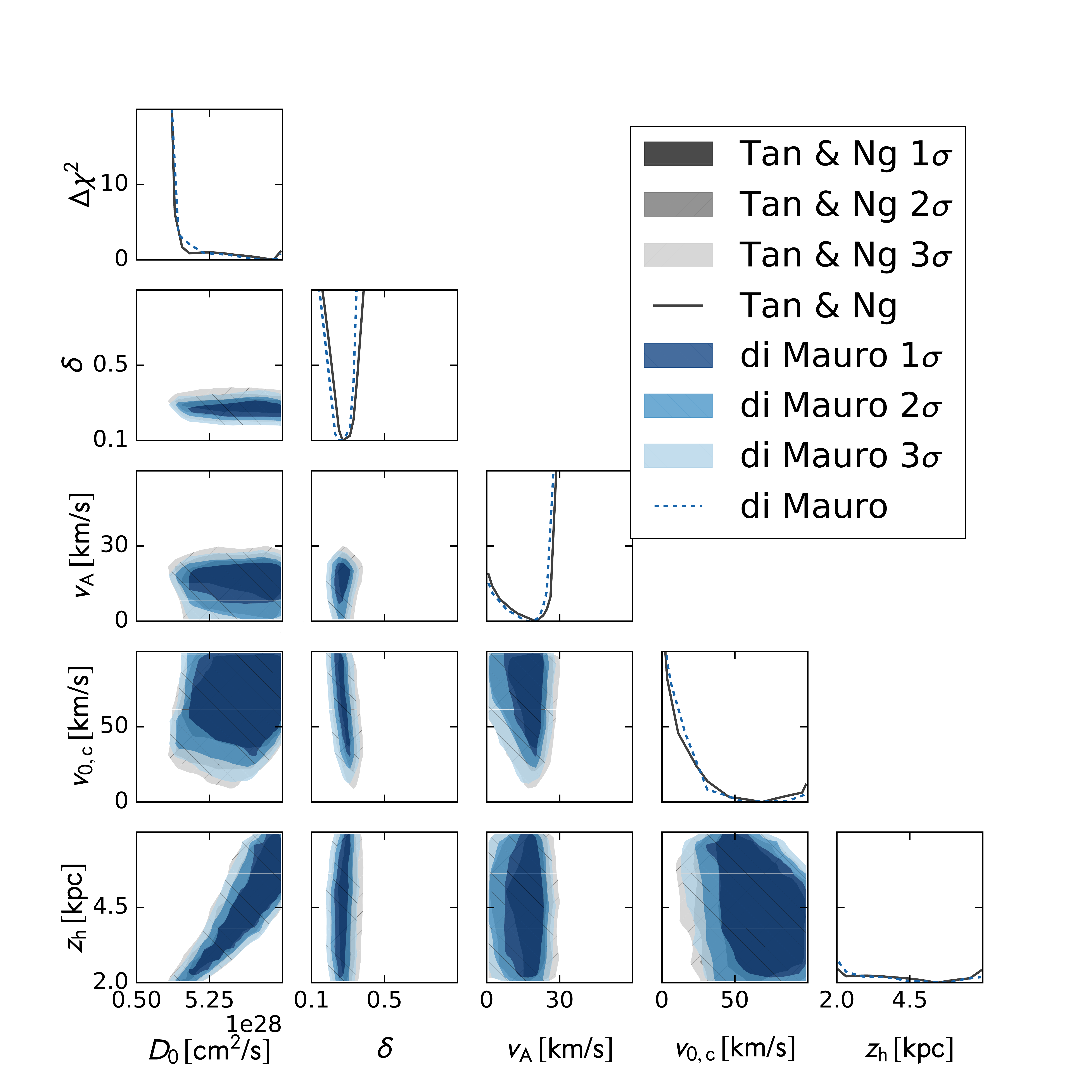}
  \caption{Comparison of the baseline fit with the fit using the antiproton 
                          production cross section from di Mauro et al.}
	\label{fig::comparison_diMauro_CS}
	\end{minipage}
\end{figure}

As mentioned above,  in order to get the antiproton  $(\bar{p})$ production cross section for arbitrary projectiles and targets, 
denoted by subscripts $P$ and $T$, respectively, we need a scaling from  the $pp$ collision cross section. 
A common approximation is to treat the projectile as a flux of $A_P$ protons with momenta $p_P/A_P$.
The target is instead scaled according to the semiclassical model where the volume scales as $A_T$
and the area as $A_T^{2/3}$.
For the  production of antineutrons $(\bar{n})$, which subsequently  decay into antiprotons 
and contribute directly to the flux, we assume a scaling of $1.3$. 
Comprehensively, we get
\begin{eqnarray}
  \label{eqn::AntiprotonCS_scaleing}
  \frac{\diff \sigma_{P, T}^{({\bar{p}})}  (p_P, p_{\bar{p}})}{\diff p_{\bar{p}}}  &=&  A_P \, A_T^{2/3} \,  \frac{\diff \sigma_{p, p}^{(\bar{p})}  \left(\frac{p_P}{A_P}, p_{\bar{p}}\right)}{\diff p_{\bar{p}}} \\
  \frac{\diff \sigma_{P, T}^{({\bar{n}})}  (p_P, p_{\bar{n}})}{\diff p_{\bar{n}}} &=& 1.3 \, \frac{\diff \sigma_{P, T}^{(\bar{p})}  (p_P, p_{\bar{n}})}{\diff p_{\bar{n}}}.
\end{eqnarray}
The dominant contribution of the antiproton flux comes from $pp$ collisions (cf. \figref{5GV_PHePbar_pbar}).
The contribution from nonproton projectiles and/or targets plays a subdominant role, 
since both the interstellar helium gas contribution and the CR abundance of helium amount to only roughly $10\%$. 
The main uncertainty arising from the scaling is thus related to
the antineutron cross section, since the antineutrons are produced directly in the $pp$ collisions.
The scaling described above gives similar results to the one implemented as default in \textsc{Galprop}.
\figref{5GV_PHePbar_pbar} shows the comparison between the total antiproton flux for our best fit model 
in the main fit framework
using the default cross section and the one derived in \cite{Mauro_Antiproton_Cross_Section_2014} (their Eq.~(13)).
We can see that overall  the cross section from \cite{Mauro_Antiproton_Cross_Section_2014} predicts
a lower normalization of the antiproton flux by about $\sim$20\%. Also the shape is slightly different
with a mild hardening of the flux starting at about $\sim$ 20$\,$GV. 
In the plot, for completeness, we also show the contribution from tertiary antiprotons,
and the separate contributions from proton collisions and collisions involving helium. 
Finally,  we also tested a different scaling available in  \textsc{Galprop} due to Simon et al. \cite{Simon_Antiproton_CS_Scaling_1998}
applied to the two cross sections, and we find in both cases that this introduces a $\sim$ 5\% variation
with respect to the flux with the default scaling.

Intriguingly, as  seen in \figref{5GV_PHePbar_pbar} , the flux derived from the cross section from   \cite{Mauro_Antiproton_Cross_Section_2014} 
seems to slightly better fit the observed antiproton spectrum.
We thus repeat the fit using the new cross section (diMauro). \figref{5GV_PHePbar_diMauro_CS} shows the results using the parametrization from their Eq.~(13). As expected the high-energy part fits better than in the (main) fit, but the shape of  the low-energy tail does not exactly match the measurements leaving a similar amount of systematics in the residuals. As expected the lower normalization of the cross section compared to \textsc{Galprop} is compensated by a slightly lower value of $\delta$ which  drops to $0.27$, which can be seen from \figref{comparison_diMauro_CS}. The other parameters are not changed. 
We also test the cross section from their Eq.~(12), which leads to a similar result.


\subsection{Frequentist vs. Bayesian interpretation}
\label{subsec::Bayes}

The results of this analysis are interpreted in the frequentist approach, whereas previous analyses where mostly done in a Bayesian framework. 
We thus compare the two approaches  for the  main fit case. 
In the Bayesian case we derive constraints from the posterior distribution, as opposed to the frequentist case where we use only the likelihood function.
The Bayesian posterior is interpreted as probability distribution once the priors in model parameters are specified. 
In our case, prior ranges are as specified in \tabref{Parameters},
and they are linear in all the parameters.
Two-dimensional posteriors for two given parameters are derived marginalizing (i.e., integrating) the full posterior over the remaining parameters.
In practice, marginalized posteriors are a natural output of the Monte Carlo based  scanning technique, 
and integrals do not need to be performed explicitly.
Bayesian contours  are then derived integrating the marginalized posterior up to the specified  confidence level.
In \figref{5GV_frequentist_comp_bayesian_triangle} we show the  triangle plot for a selected set of parameters
and compare the $1\sigma$ to $3\sigma$ frequentist contours with the equivalent Bayesian contours.
It can be seen that the two approaches give compatible results, with the frequentist case being slightly more conservative.
In fact the two approaches are expected to give compatible results in the limit in which the data are 
 constraining enough, and the effect of the priors start to be subdominant. 
The above results indicate indeed that the results are data driven rather than prior driven, and thus robust.

\begin{figure}[t!]
	\centering
	\includegraphics[width=0.48\textwidth]
    {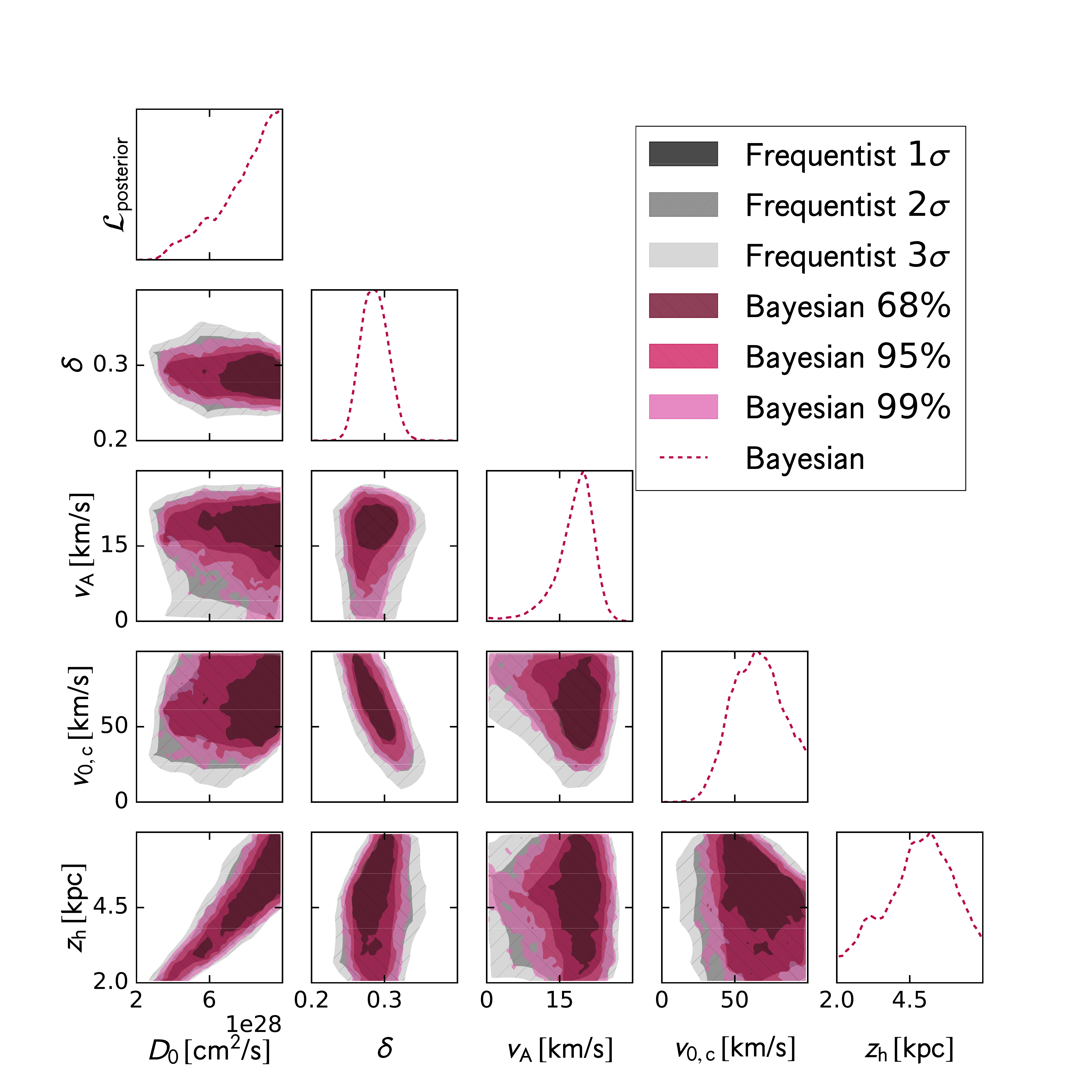}
  \caption{Comparison of the baseline fit (main) with the Bayesian interpretation.}
  \label{fig::5GV_frequentist_comp_bayesian_triangle}
\end{figure}


\begin{figure}[t!]
	\centering
	\begin{subfigure}[b]{0.48\textwidth}
	  \includegraphics[width=\textwidth]
	    {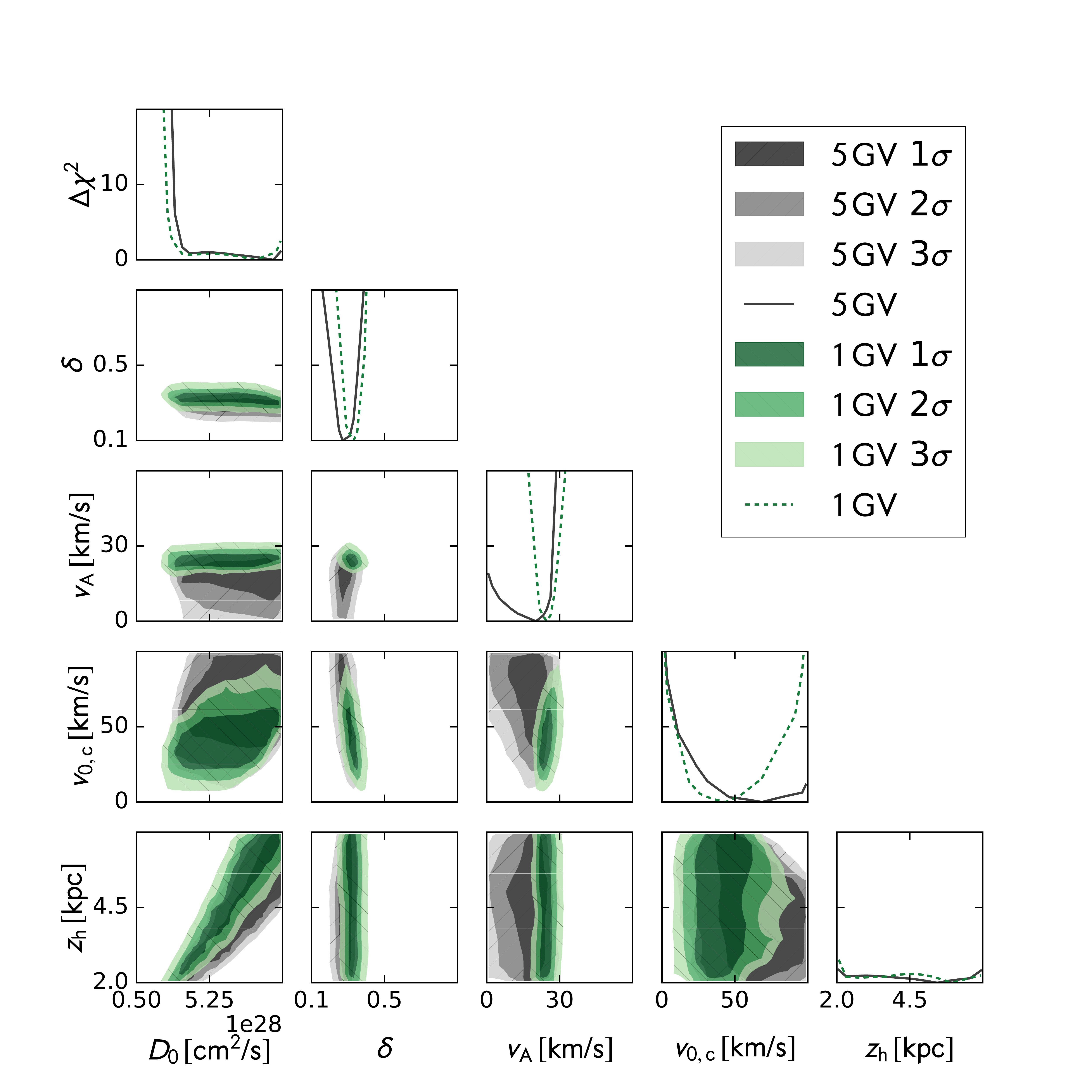}
  	\caption{Propagation parameters.}
	  \label{fig::5GV_comp_1GV_triangle}
	\end{subfigure}
	\begin{subfigure}[b]{0.48\textwidth}
	  \includegraphics[width=\textwidth]
	    {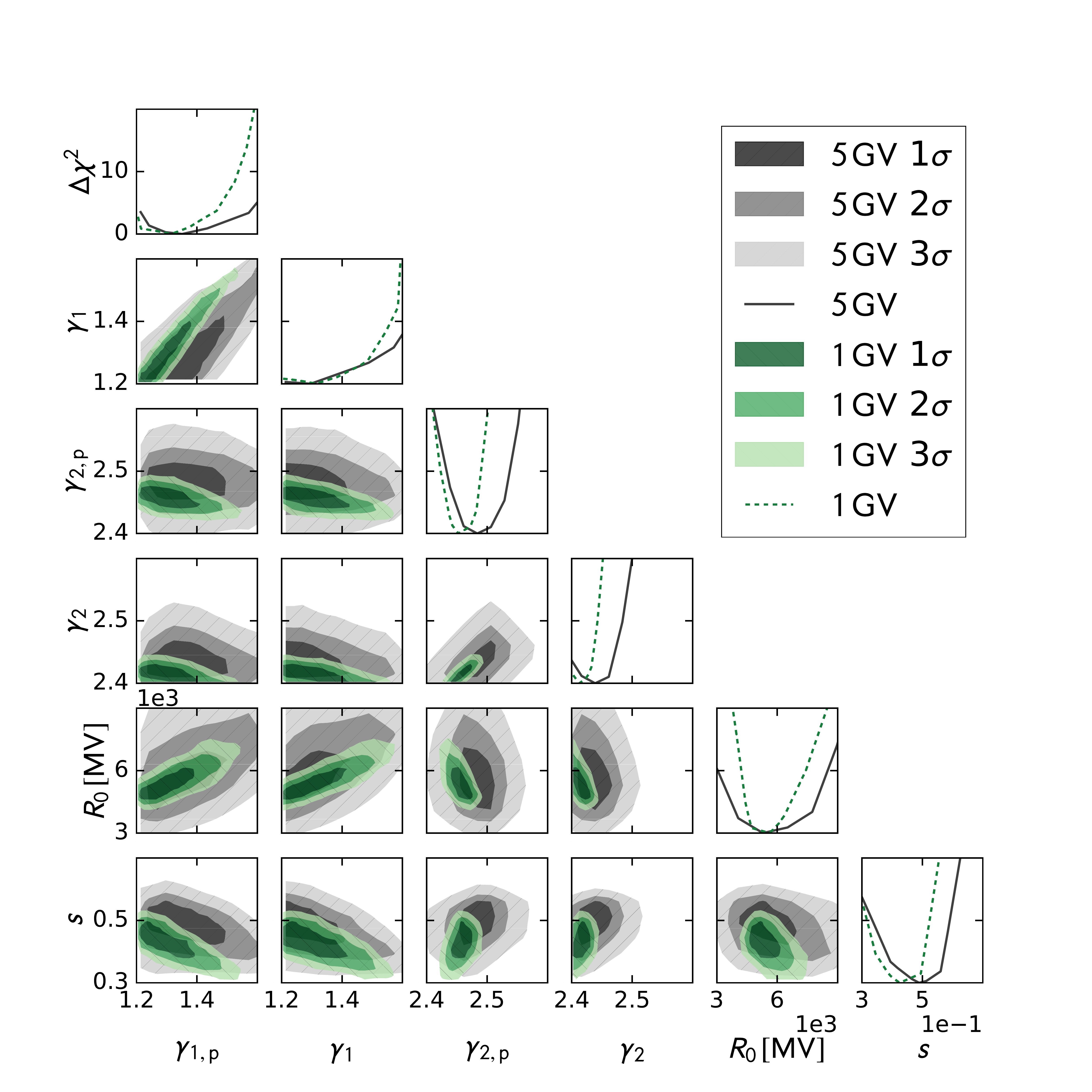}
    \caption{Injection parameters}
	  \label{fig::5GV_comp_1GV_triangle_inj}
	\end{subfigure}
	\caption{Comparison of the baseline fit (main) to a fit including data down to $1\,$GV (1GV)for 
	  (a) propagation and (b) injection parameters.}
  \label{fig::Comparison_triangles_1GV}
\end{figure}

\subsection{Fit down to 1$\,$GV}
\label{subsec::1GV}

\begin{figure}[t!]
	\centering
	\begin{subfigure}[b]{0.3\textwidth}
	  \includegraphics[width=\textwidth]{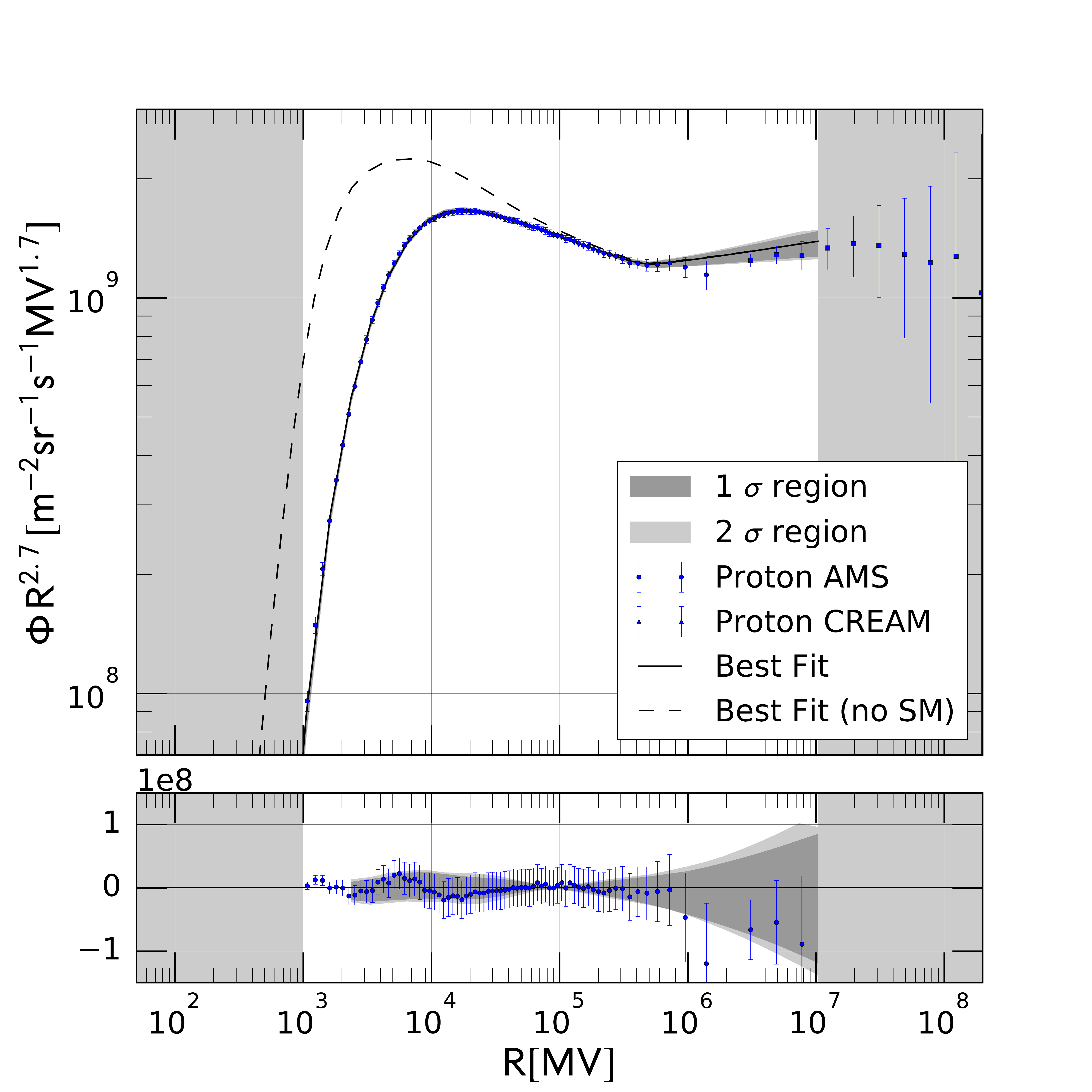}
  	\caption{Proton}
	  \label{fig::1GV_PHePbar_bestFit_P}
	\end{subfigure}
	\begin{subfigure}[b]{0.3\textwidth}
	  \includegraphics[width=\textwidth]{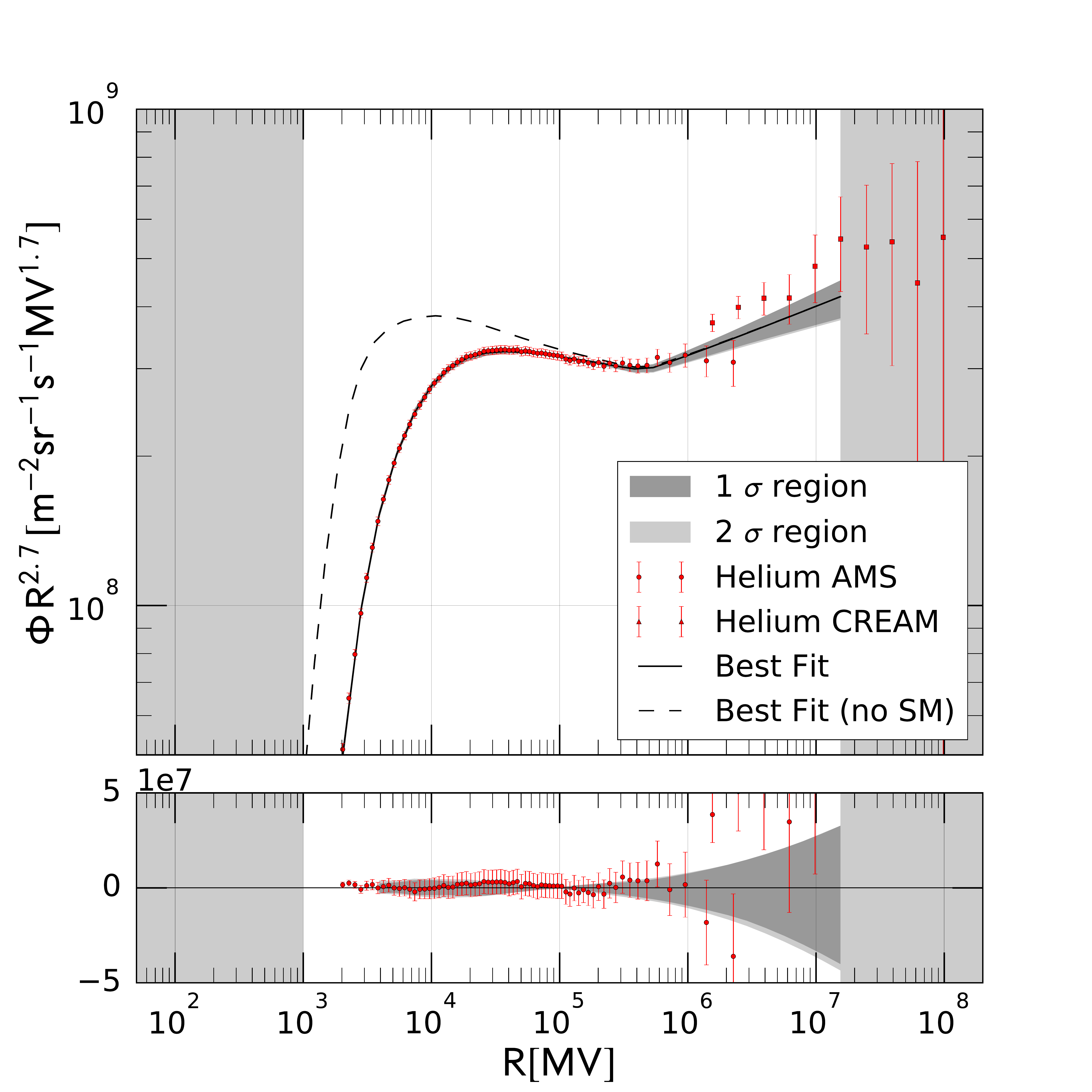}
    \caption{Helium}
	  \label{fig::1GV_PHePbar_bestFit_He}
	\end{subfigure}
	\begin{subfigure}[b]{0.3\textwidth}
	  \includegraphics[width=\textwidth]{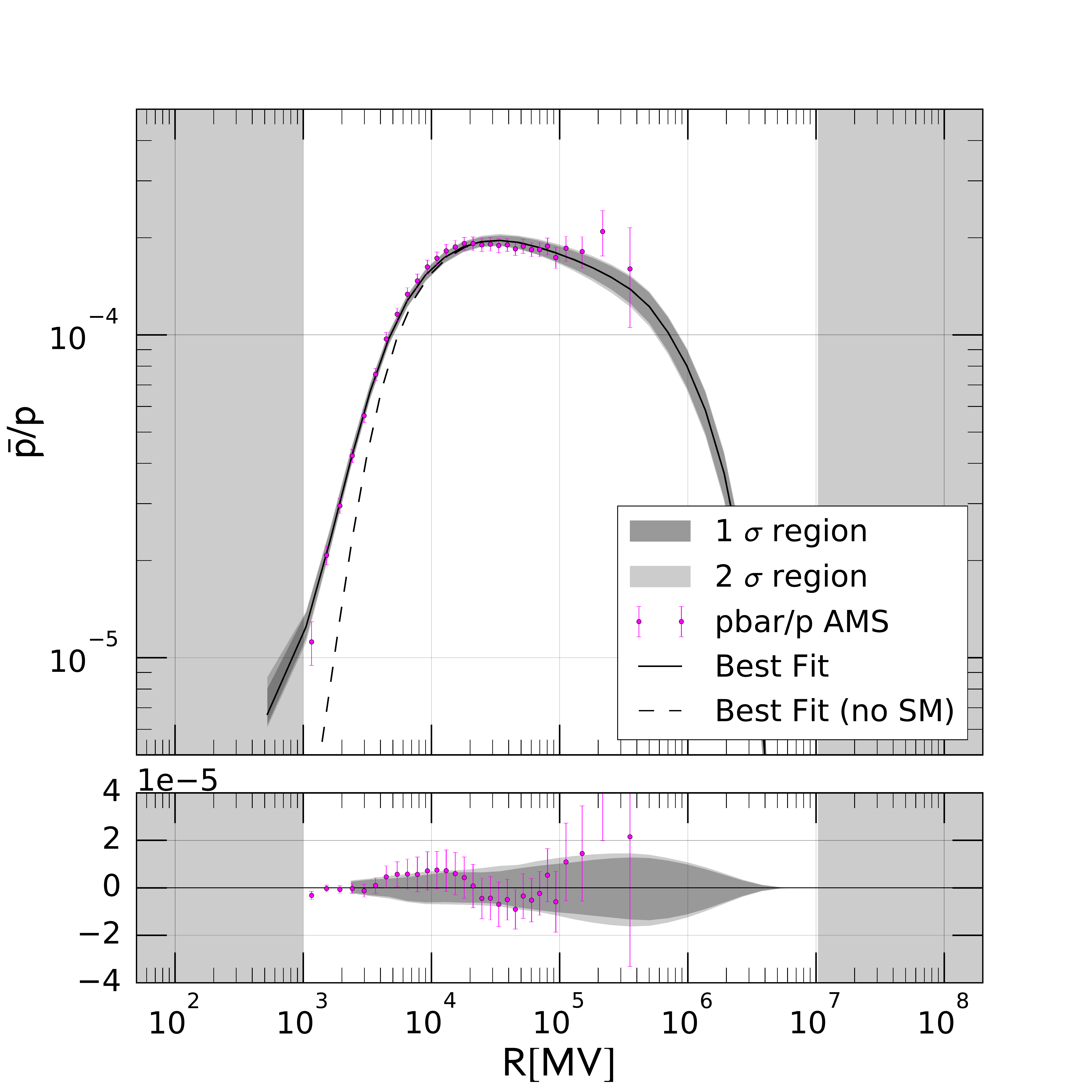}
    \caption{$\bar{p}/p$}
	  \label{fig::1GV_PHePbar_bestFit_Pbar}
	\end{subfigure}
	\caption{Comparison between data and best-fit model for the global analysis including proton, helium, and antiprotons down to 1$\,$GV (1GV).}
	\label{fig::1GV_PHePbar_bestFit}
\end{figure}

As explained in \secref{methods} we limit our fit range to $R> 5\,$GV to reduce the effects of the solar modulation. Thus we avoid the rigidity range indicating rigidity and charge sign dependence of the solar modulation potential.
In this section we investigate the effect of extending the fit range down to $1\,$GV (labeled (1GV)).
The results of the fit are shown in \figref{Comparison_triangles_1GV} and \figref{1GV_PHePbar_bestFit}.
\figref{1GV_PHePbar_bestFit} shows that a good fit is achieved with flat residuals all
over the fitted energy range. The $\chi^2/$NDF has a value of 70.3/175.
From  \figref{5GV_comp_1GV_triangle} it can be seen that the two fits give consistent results
at the level of a bit more than $2\sigma$. The slight shift of about $0.04$ in the value of $\delta$ from $0.28^{+0.03}_{-0.01}$
to $0.32^{+0.03}_{-0.02}$ can be considered as an estimate of the systematic
error on this parameter.
Regarding reacceleration and convection, adding data down to $1\,$GV contributes to basically break their degeneracy
providing a strong constraint on $v_{\rm A}=25.0    ^{ +0.92    } _{ -2.30     }\,$km/s and a lower value of $v_{0,c}=44.4^{+12.2}_{-19.8}\,$km/s.
At the same time tight constraints on the index below the break $\gamma_1= 1.32    ^{ +0.06    } _{ -\I{0.12}}$,
$\gamma_{1,p}=1.32    ^{ +0.05    } _{ -\I{0.12} }$ and  on the break itself  $R_0=5.52    ^{ +0.33    } _{ -0.83     }\,$GV appear.
The latter value suspiciously coincides with  the rigidity below which the constant solar modulation potential approximation should start to fail.
For this reason it is unclear if the presence of the break is indeed physical or if it is a way for the fit to compensate for the nonprecise solar modulation modeling. At the same time it is equally unclear if the resulting values of $v_{\rm A}$ and $v_{0,c}$ are robust or  are biased by the  possibly incorrect solar modulation.
As mentioned also in \subsecref{main}, to settle the issue a more careful study of the solar modulation effect
will be necessary, complemented by the use of time series of CR data.


\subsection{Fit without convection}
\label{subsec::convection}

\begin{figure}[t!]
	\centering
	\begin{minipage}{.45\textwidth}
  \includegraphics[width=1.\textwidth]
    {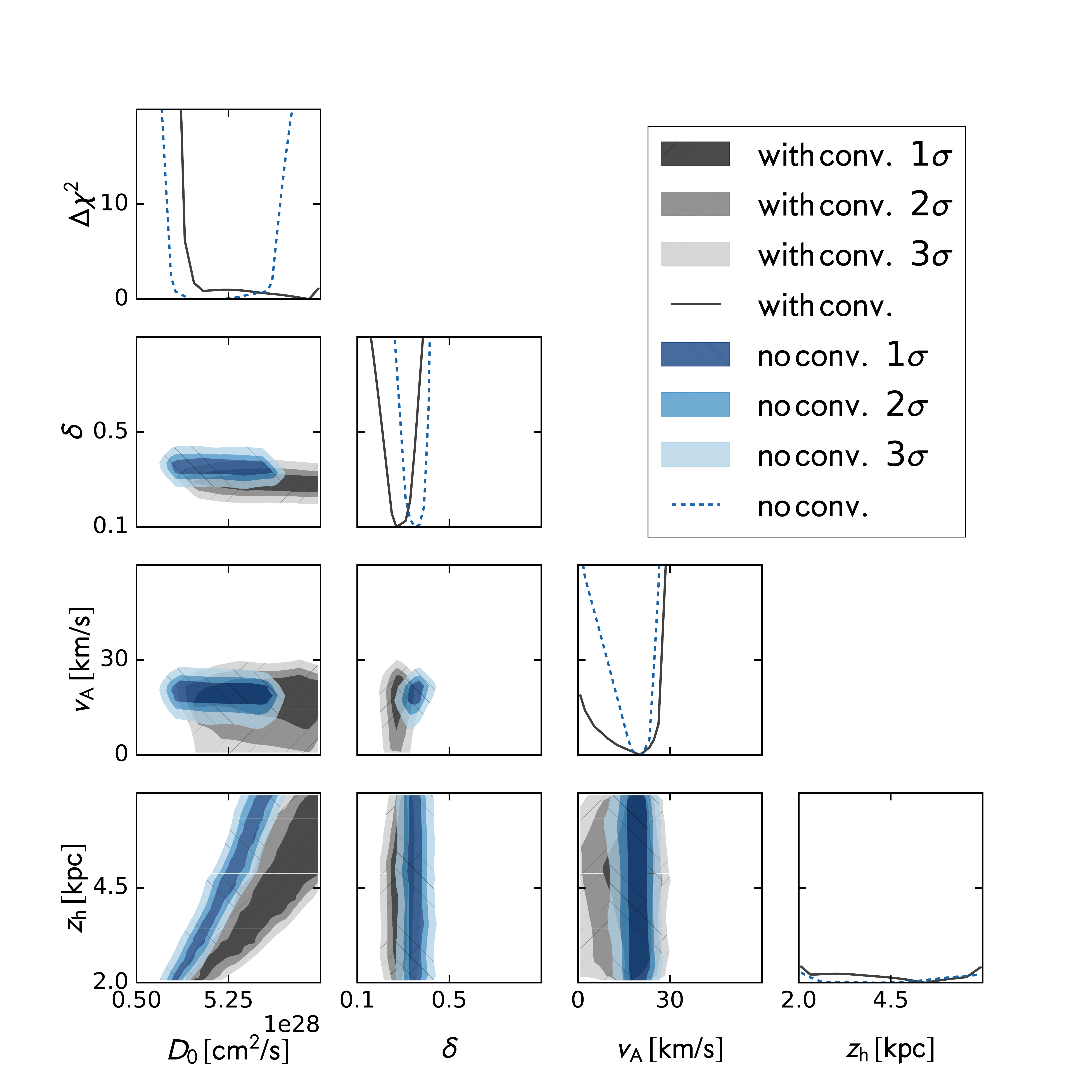}
  \caption{Comparison of the propagation parameters from the baseline analysis (main) to a diffusion model without convection (noVc-5GV).}
  \label{fig::5GV_PHePbar_NoConvection}
	\end{minipage}
	\hspace{.03\textwidth}
  \begin{minipage}{.45\textwidth}
	\centering
	\includegraphics[width=1.\textwidth]
    {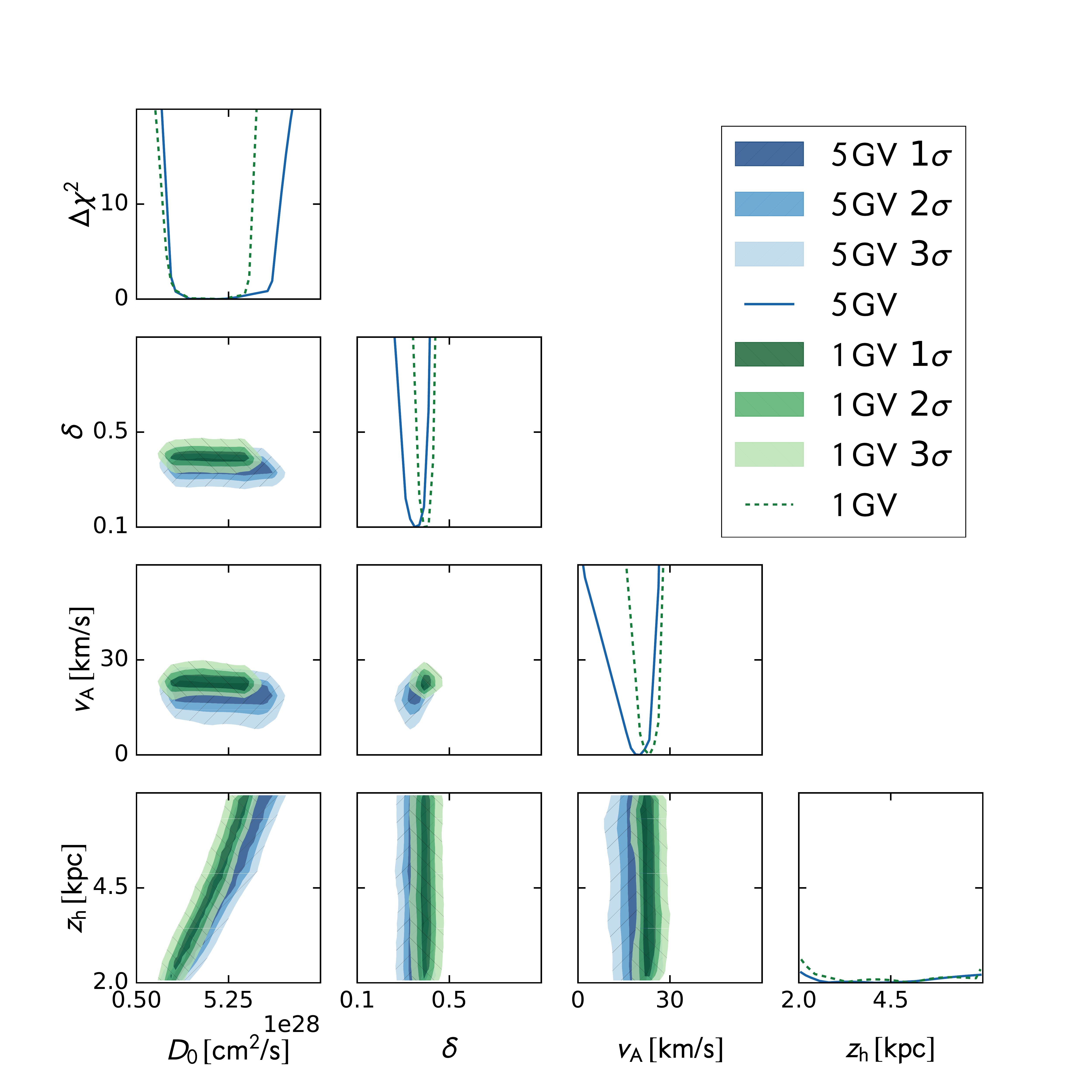}
  \caption{Comparison of the propagation parameters in the case of no convection for a 
                              rigidity cutoff at 1$\,$GV (noVc-1GV) and 5$\,$GV (noVc-5GV).}
  \label{fig::1GV_PHePbar_NoConvection}
	\end{minipage}
\end{figure}

\begin{figure}[t!]
	\centering
	\begin{subfigure}[b]{0.3\textwidth}
	  \includegraphics[width=\textwidth]{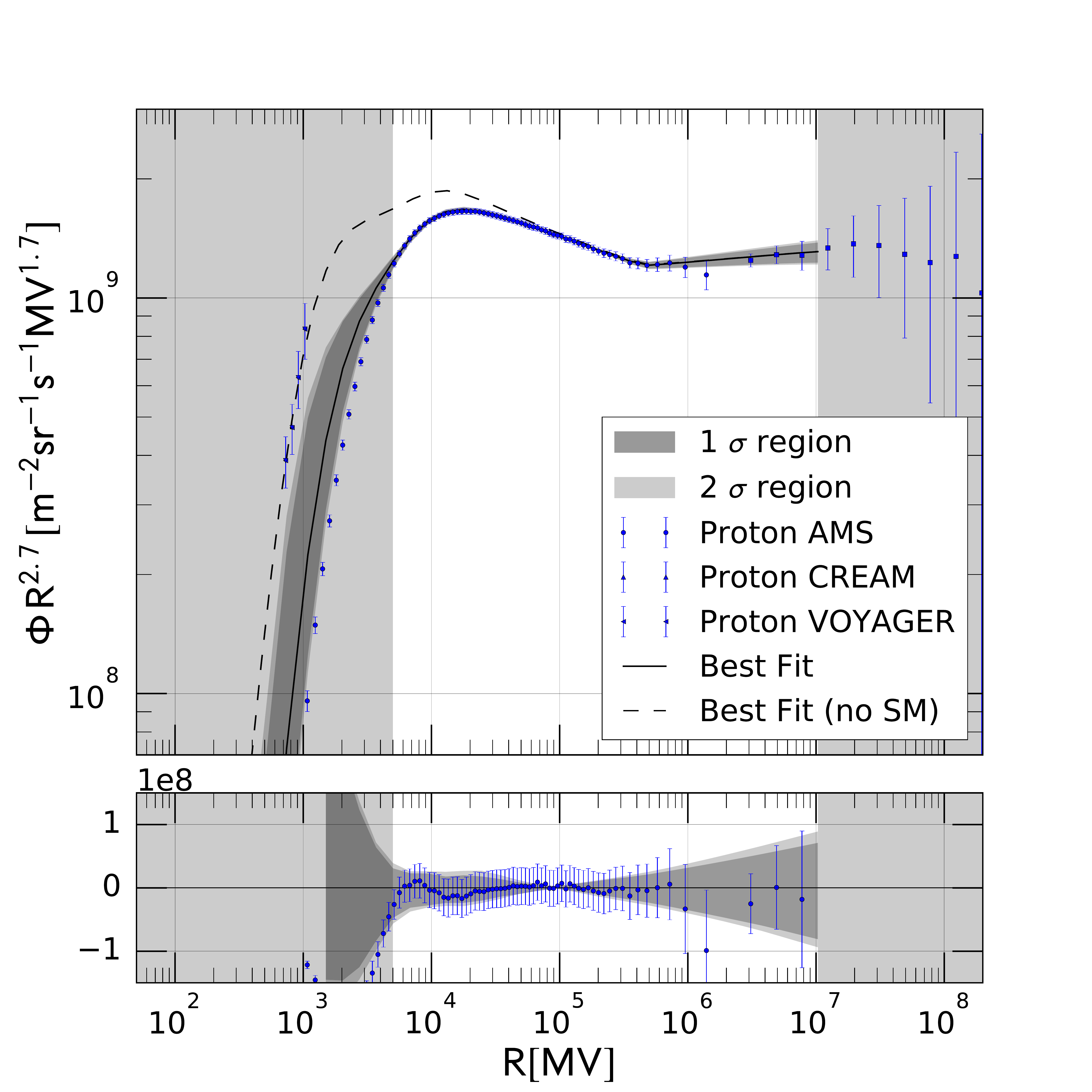}
  	\caption{Proton}
	  \label{fig::noConv_PHePbar_bestFit_P}
	\end{subfigure}
	\begin{subfigure}[b]{0.3\textwidth}
	  \includegraphics[width=\textwidth]{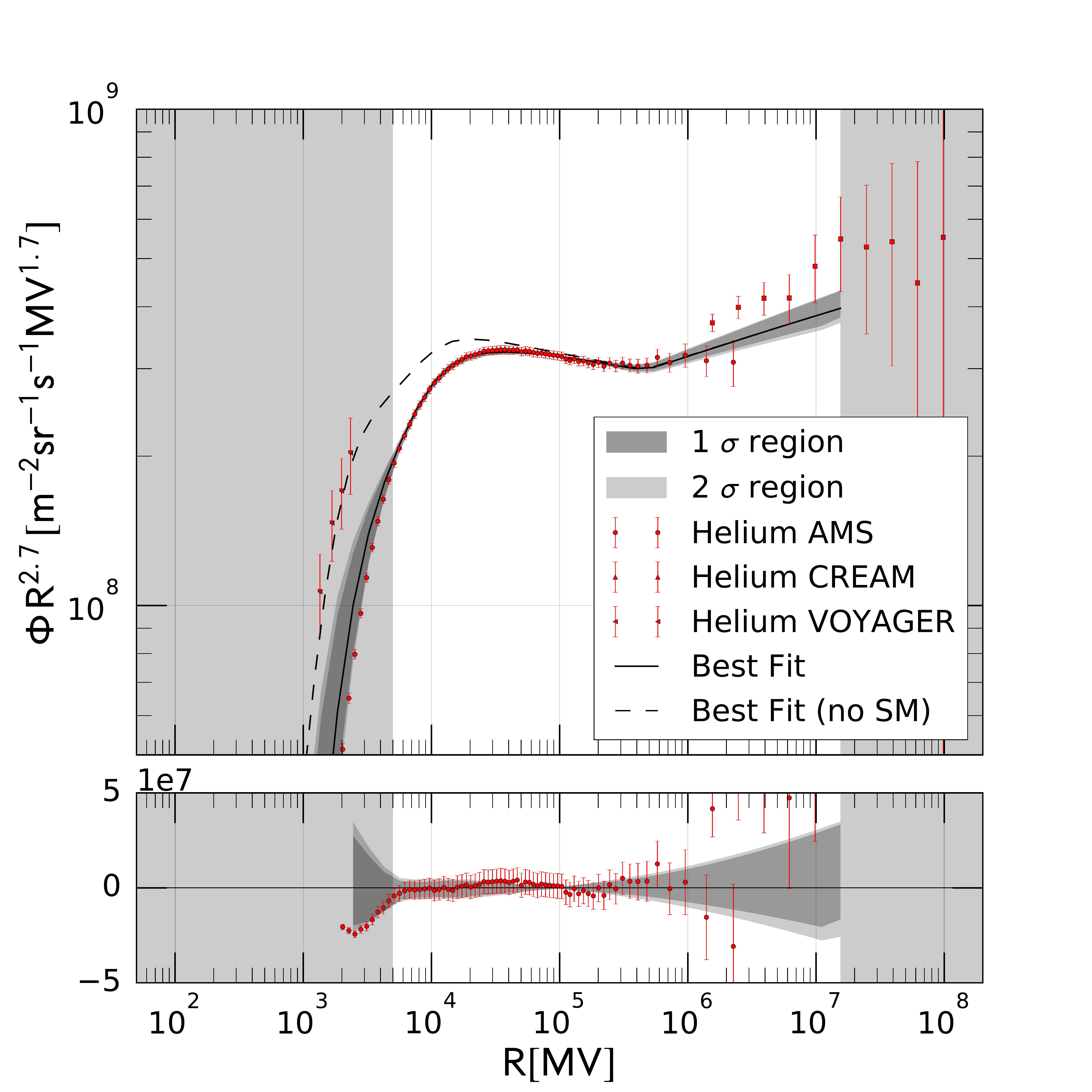}
    \caption{Helium}
	  \label{fig::noConv_PHePbar_bestFit_He}
	\end{subfigure}
	\begin{subfigure}[b]{0.3\textwidth}
	  \includegraphics[width=\textwidth]{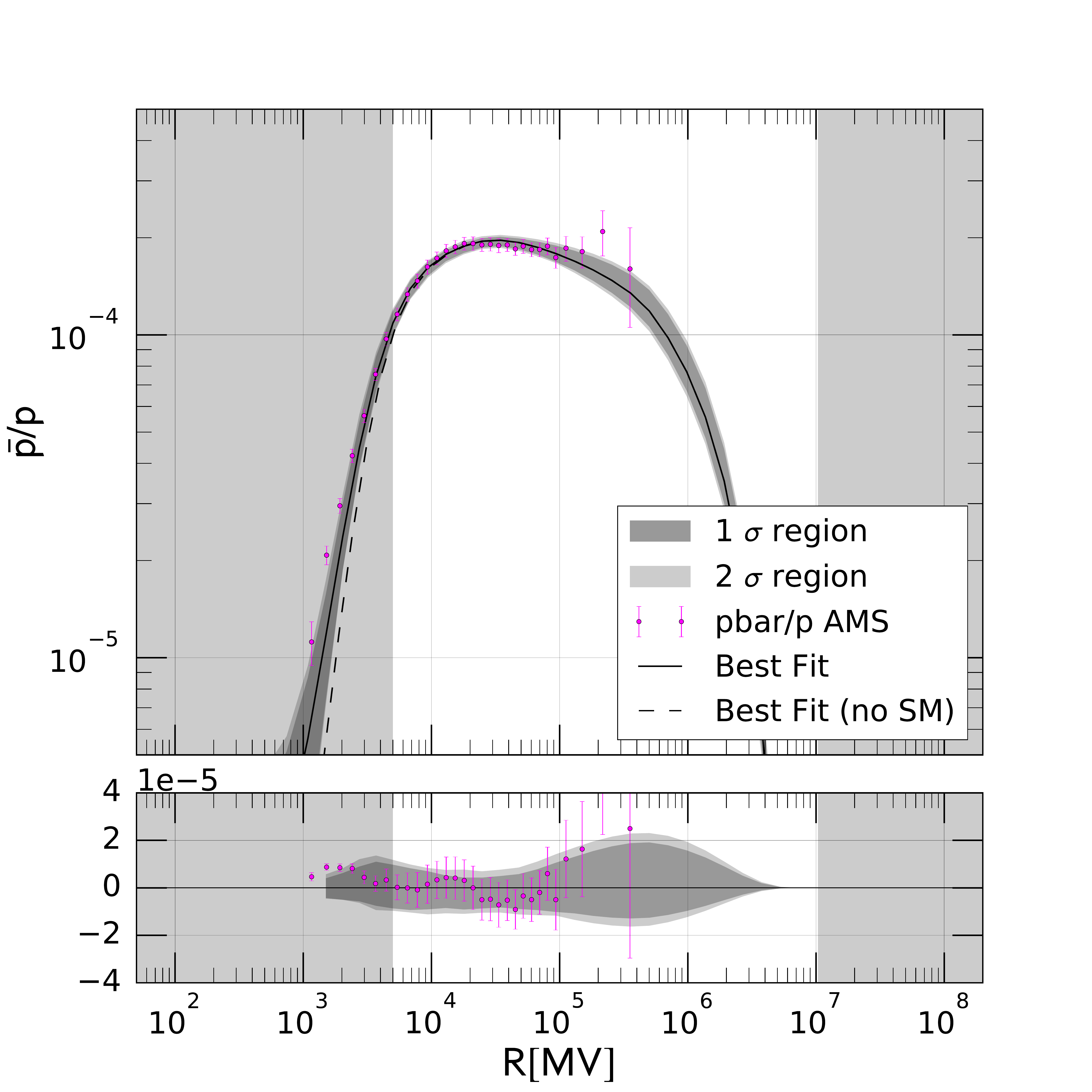}
    \caption{$\bar{p}/p$}
	  \label{fig::noConv_PHePbar_bestFit_Pbar}
	\end{subfigure}
	\caption{Comparison between data and best-fit model for a diffusion model without convection and rigidity cut at 5$\,$GV (noVc-5GV).}
	\label{fig::noConv_PHePbar_bestFit}
\end{figure}

\begin{figure}[t!]
	\centering
	\begin{subfigure}[b]{0.3\textwidth}
	  \includegraphics[width=\textwidth]{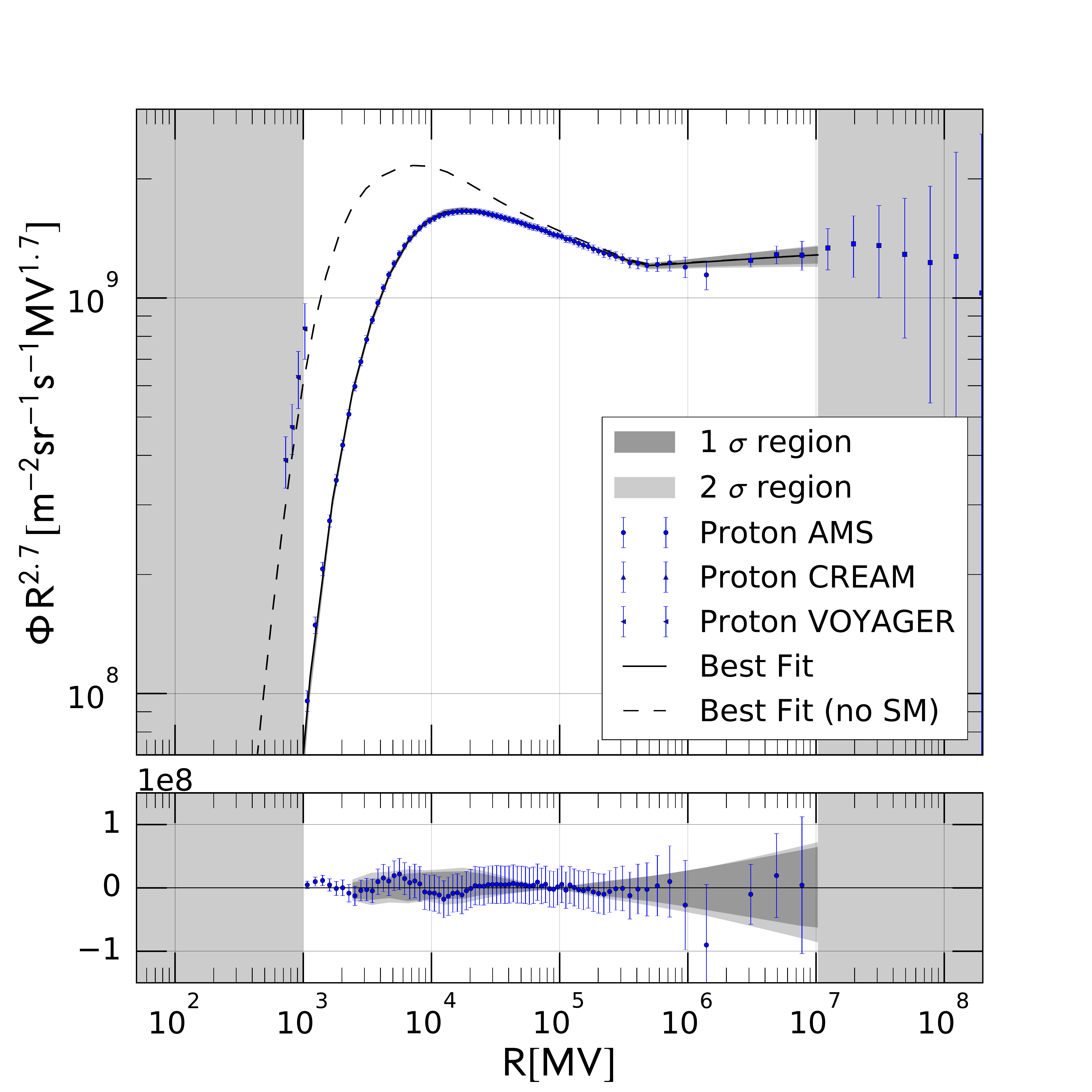}
  	\caption{Proton}
	  \label{fig::noConv1GV_PHePbar_bestFit_P}
	\end{subfigure}
	\begin{subfigure}[b]{0.3\textwidth}
	  \includegraphics[width=\textwidth]{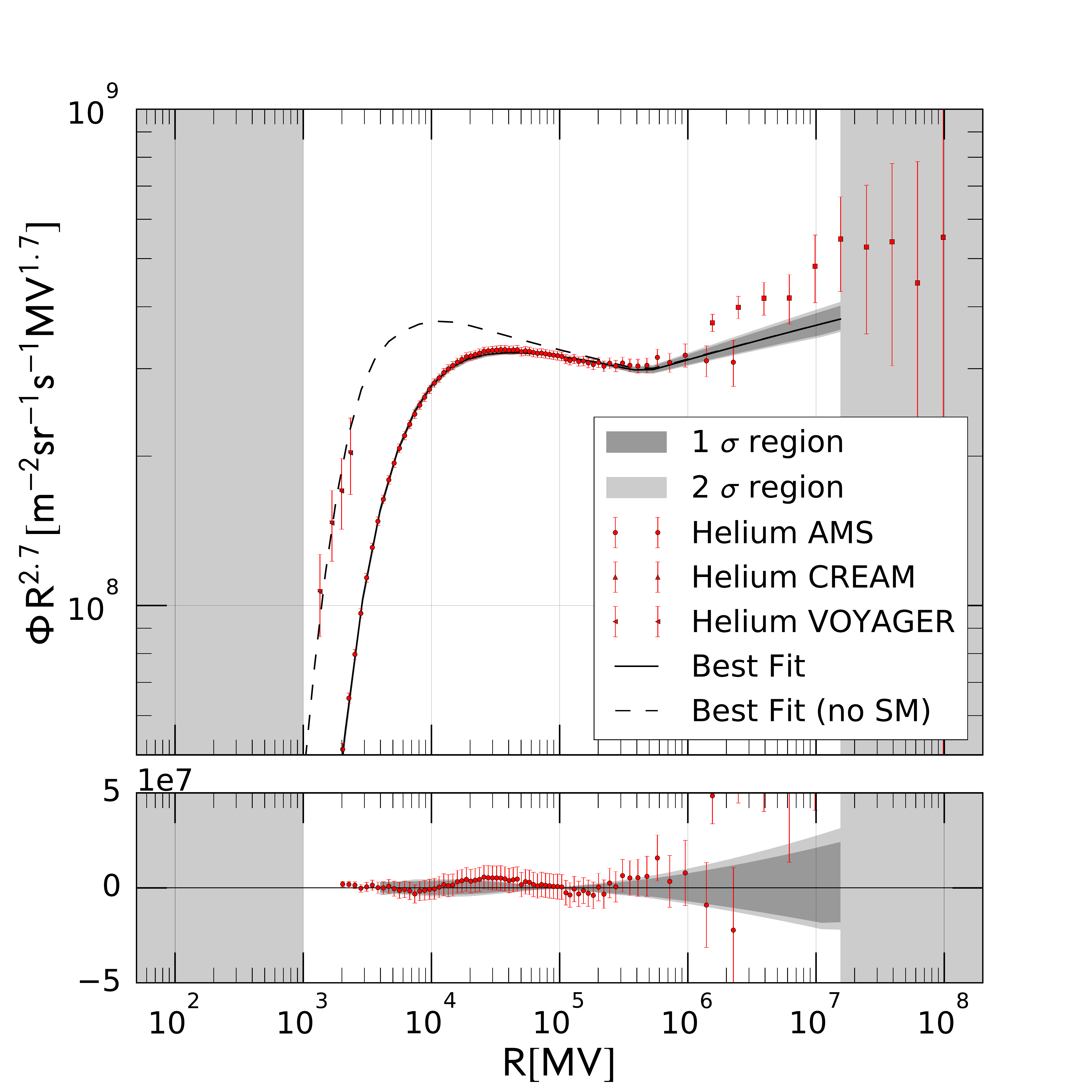}
    \caption{Helium}
	  \label{fig::noConv1GV_PHePbar_bestFit_He}
	\end{subfigure}
	\begin{subfigure}[b]{0.3\textwidth}
	  \includegraphics[width=\textwidth]{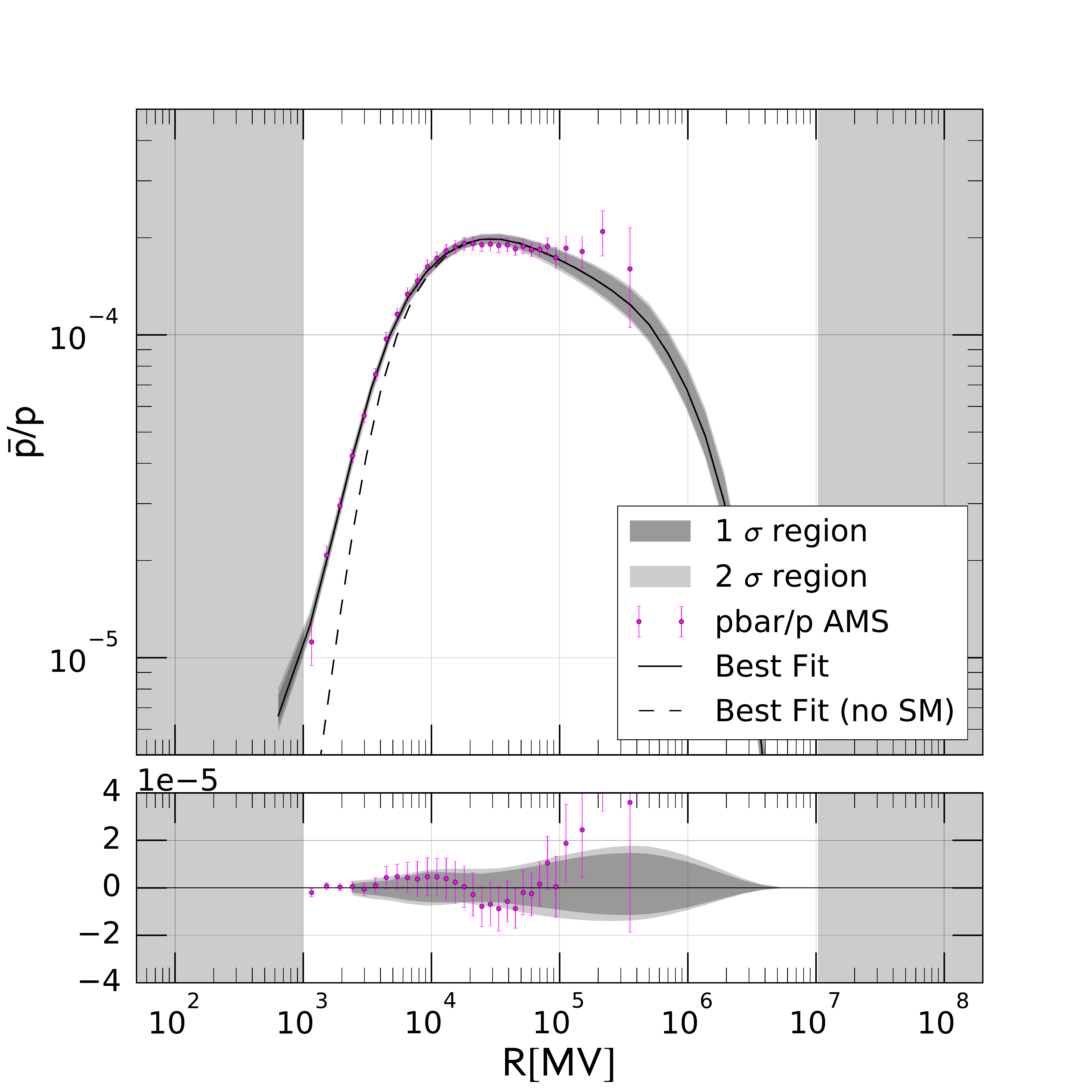}
    \caption{$\bar{p}/p$}
	  \label{fig::noConv1GV_PHePbar_bestFit_Pbar}
	\end{subfigure}
	\caption{Comparison between data and best-fit model for a diffusion model without convection and with data down to 1$\,$GV (noVc-1GV).}
	\label{fig::noConv1GV_PHePbar_bestFit}
\end{figure}

As last cross-check we investigate the necessity of convection. The baseline fit allows  convection velocities 
and finds a tendency toward large values $\agt 50\,$km/s,  while the fit including data down to $1\,$GV prefers low values.
Therefore, we also test a diffusion model without convection (labeled (noVc-5GV)) and compare it to our baseline fit. 
This also allows a more direct comparison with previous works where convection was not considered \cite{Johannesson_CR_Propagation_2016, Trotta_CR_Propagation_2011}.
Results are shown in \figref{5GV_PHePbar_NoConvection} and \figref{noConv_PHePbar_bestFit}  .
The residuals in \figref{noConv_PHePbar_bestFit} show a flat behavior. The resulting $\chi^2/$NDF is $48.7/146$.
On the other hand the amount of solar modulation $340^{+45}_{-125}\,$MV appears to be too low.
\figref{5GV_PHePbar_NoConvection} shows the comparison of the remaining propagation parameters to the baseline fit. 
The model without convection prefers slightly higher values for $\delta$, slightly lower values for $D_0$,
and a well constrained value of $v_{\rm A}$, so that,
overall the results seem more in agreement with the baseline fit extending down to 1$\,$GV, rather than the main case down to 5$\,$GV.
Given also the low value for the solar modulation in the case (noVc-5GV) we thus also tested the case 
of data down to 1$\,$GV (labeled noVc-1GV). 
In this case the $\chi^2/$NDF increases to $82.2/176$, but we achieve a more reasonable value for the solar modulation potential of $640\pm 20\,$MV. 
The propagation parameters are not much affected by the data below $5\,$GV as can be seen from \figref{1GV_PHePbar_NoConvection}.
 \figref{noConv1GV_PHePbar_bestFit} shows that residuals are reasonably flat also for this fit.

\begin{table}[H]
  \caption{Best fit values and 1$\sigma$ errors for the various fits. 
              If the error coincides with the upper or lower bound of the prior range
              the error value is given in italic. }
  \label{tab::BestFit}
  \centering
  \begin{tabular}{l l r r r r r r r r r }
  \hline \hline
  \multicolumn{2}{l}{\textbf{Fit parameters}} 
  & \multicolumn{1}{c}{\text{(uni-PHe)}}   
  & \multicolumn{1}{c}{\text{(uni-PHePbar)}}   
  & \multicolumn{1}{c}{\text{(P)}}   
  & \multicolumn{1}{c}{\text{(PHe)}}   
  & \multicolumn{1}{c}{\text{(main)}} 
  & \multicolumn{1}{c}{\text{(diMauro)}} 
  & \multicolumn{1}{c}{\text{(1GV)}} 
  & \multicolumn{1}{c}{\text{(noVc-1GV)}} 
  & \multicolumn{1}{c}{\text{(noVc-5GV)}}  \\ \hline \hline
  $\gamma_{1,p}       $  & $        $ 
                                              & -                                                
                                              & -                                                
                                              & $1.52    ^{ +0.21    } _{ -\I{0.32} }    $       
                                              & $1.27    ^{ +0.11    } _{ -\I{0.07} }    $       
                                              & $1.36    ^{ +0.07    } _{ -0.10     }    $       
                                              & $1.38    ^{ +0.07    } _{ -0.10     }    $       
                                              & $1.32    ^{ +0.05    } _{ -\I{0.12} }    $       
                                              & $1.61    ^{ +0.06    } _{ -0.10     }    $       
                                              & $1.76    ^{ +0.07    } _{ -0.04     }    $ \\    
  $\gamma_{2,p}       $  & $        $ 
                                              & -                                                
                                              & -                                                
                                              & $2.52    ^{ +0.12    } _{ -0.45     }    $       
                                              & $2.069   ^{ +0.098   } _{ -\I{0.069}}    $       
                                              & $2.493   ^{ +0.010   } _{ -0.026    }    $       
                                              & $2.499   ^{ +0.026   } _{ -0.014    }    $       
                                              & $2.455   ^{ +0.014   } _{ -0.007    }    $       
                                              & $2.421   ^{ +0.010   } _{ -0.014    }    $       
                                              & $2.454   ^{ +0.026   } _{ -0.014    }    $ \\    
  $\gamma_{1}         $  & $        $ 
                                              & $1.92    ^{ +0.08    } _{ -0.14     }    $       
                                              & $1.50    ^{ +0.07    } _{ -0.12     }    $       
                                              & -                                                
                                              & $1.53    ^{ +0.24    } _{ -0.11     }    $       
                                              & $1.29    ^{ +0.04    } _{ -\I{0.09} }    $       
                                              & $1.26    ^{ +0.10    } _{ -\I{0.06} }    $       
                                              & $1.32    ^{ +0.06    } _{ -\I{0.12} }    $       
                                              & $1.65    ^{ +0.07    } _{ -0.11     }    $       
                                              & $1.70    ^{ +0.06    } _{ -0.07     }    $ \\    
  $\gamma_{2}         $  & $        $ 
                                              & $2.582   ^{ +0.010   } _{ -0.034    }    $       
                                              & $2.404   ^{ +0.006   } _{ -0.022    }    $       
                                              & -                                                
                                              & $2.003   ^{ +0.094   } _{ -\I{0.003}}    $       
                                              & $2.440   ^{ +0.006   } _{ -0.018    }    $       
                                              & $2.451   ^{ +0.018   } _{ -0.010    }    $       
                                              & $2.412   ^{ +0.012   } _{ -0.006    }    $       
                                              & $2.381   ^{ +0.010   } _{ -0.010    }    $       
                                              & $2.407   ^{ +0.022   } _{ -0.014    }    $ \\    
  $R_0                $  & [GV]$    $ 
                                              & $8.16    ^{ +1.22    } _{ -1.54     }    $       
                                              & $8.79    ^{ +1.17    } _{ -1.55     }    $       
                                              & $4.38    ^{ +3.23    } _{ -1.54     }    $       
                                              & $10.5    ^{ +1.40    } _{ -1.59     }    $       
                                              & $5.54    ^{ +0.76    } _{ -0.54     }    $       
                                              & $5.44    ^{ +0.54    } _{ -0.54     }    $       
                                              & $5.52    ^{ +0.33    } _{ -0.83     }    $       
                                              & $7.01    ^{ +0.98    } _{ -0.54     }    $       
                                              & $8.63    ^{ +0.98    } _{ -0.76     }    $ \\    
  $s                  $  & $        $ 
                                              & $0.32    ^{ +0.08    } _{ -0.02     }    $       
                                              & $0.41    ^{ +\I{0.09}} _{ -0.07     }    $       
                                              & $0.48    ^{ +0.16    } _{ -0.31     }    $       
                                              & $0.59    ^{ +0.16    } _{ -0.04     }    $       
                                              & $0.50    ^{ +0.02    } _{ -0.04     }    $       
                                              & $0.50    ^{ +0.05    } _{ -0.03     }    $       
                                              & $0.43    ^{ +0.04    } _{ -0.03     }    $       
                                              & $0.31    ^{ +0.03    } _{ -0.03     }    $       
                                              & $0.32    ^{ +0.04    } _{ -0.05     }    $ \\    
  $\delta             $  & $        $ 
                                              & $0.16    ^{ +0.03    } _{ -0.02     }    $       
                                              & $0.36    ^{ +0.04    } _{ -0.03     }    $       
                                              & $0.29    ^{ +0.46    } _{ -0.18     }    $       
                                              & $0.72    ^{ +0.01    } _{ -0.11     }    $       
                                              & $0.28    ^{ +0.03    } _{ -0.01     }    $       
                                              & $0.27    ^{ +0.02    } _{ -0.04     }    $       
                                              & $0.32    ^{ +0.03    } _{ -0.02     }    $       
                                              & $0.40    ^{ +0.01    } _{ -0.01     }    $       
                                              & $0.36    ^{ +0.02    } _{ -0.02     }    $ \\    
  $D_0                $  & [$10^{28}$ cm$^2$/s] 
                                              & $2.77    ^{ +2.95    } _{ -0.53     }    $       
                                              & $2.83    ^{ +0.90    } _{ -0.50     }    $       
                                              & $4.78    ^{ +\I{5.22}} _{ -3.49     }    $       
                                              & $5.95    ^{ +0.83    } _{ -1.37     }    $       
                                              & $9.30    ^{ +\I{0.70}} _{ -5.48     }    $       
                                              & $9.04    ^{ +\I{0.96}} _{ -3.95     }    $       
                                              & $8.19    ^{ +\I{1.81}} _{ -4.68     }    $       
                                              & $4.92    ^{ +1.12    } _{ -2.36     }    $       
                                              & $4.60    ^{ +2.71    } _{ -2.04     }    $ \\    
  $v_\mathrm{A}       $  & [km/s] $ $ 
                                              & $6.80    ^{ +1.18    } _{ -2.73     }    $       
                                              & $29.2    ^{ +2.80    } _{ -1.47     }    $       
                                              & $21.2    ^{ +\I{38.8}} _{ -\I{21.2} }    $       
                                              & $1.84    ^{ +2.36    } _{ -1.08     }    $       
                                              & $20.2    ^{ +3.26    } _{ -6.33     }    $       
                                              & $18.2    ^{ +3.15    } _{ -5.91     }    $       
                                              & $25.0    ^{ +0.92    } _{ -2.30     }    $       
                                              & $22.8    ^{ +1.46    } _{ -1.05     }    $       
                                              & $20.7    ^{ +1.14    } _{ -3.43     }    $ \\    
  $v_{0,\mathrm{c}}   $  & [km/s] $ $ 
                                              & $40.9    ^{ +\I{59.1}} _{ -5.89     }    $       
                                              & $40.2    ^{ +38.1    } _{ -25.2     }    $       
                                              & $5.82    ^{ +\I{94.2}} _{ -\I{5.82} }    $       
                                              & $87.8    ^{ +\I{12.2}} _{ -7.57     }    $       
                                              & $69.7    ^{ +22.0    } _{ -24.7     }    $       
                                              & $57.3    ^{ +41.1    } _{ -12.3     }    $       
                                              & $44.0    ^{ + 8.4    } _{ -16.5     }    $       
                                              & - $                                      $       
                                              & - $                                      $ \\    
  $z_\mathrm{h}       $  & [kpc]  $ $ 
                                              & $3.77    ^{ +\I{3.23}} _{ -\I{1.77} }    $       
                                              & $2.04    ^{ +0.40    } _{ -\I{0.04} }    $       
                                              & $4.22    ^{ +\I{2.78}} _{ -\I{2.22} }    $       
                                              & $6.55    ^{ +\I{0.45}} _{ -1.63     }    $       
                                              & $5.43    ^{ +\I{1.57}} _{ -\I{3.43} }    $       
                                              & $5.84    ^{ +\I{1.16}} _{ -\I{3.84} }    $       
                                              & $6.00    ^{ +\I{1.00}} _{ -\I{4.00} }    $       
                                              & $5.05    ^{ +\I{1.95}} _{ -\I{3.05} }    $       
                                              & $4.12    ^{ +\I{2.88}} _{ -\I{2.12} }    $ \\    
  $\phi_\mathrm{AMS}  $  & $        $ 
                                              & $300     ^{ +60      } _{ -80       }    $       
                                              & $780     ^{ +80      } _{ -40       }    $       
                                              & $620     ^{ +180     } _{ -195      }    $       
                                              & $580     ^{ +45      } _{ -115      }    $       
                                              & $400     ^{ +90      } _{ -40       }    $       
                                              & $360     ^{ +115     } _{ -45       }    $       
                                              & $700     ^{ +20      } _{ -50       }    $       
                                              & $640     ^{ +20      } _{ -20       }    $       
                                              & $340     ^{ +45      } _{ -125      }    $ \\ \hline \hline
  \end{tabular}
\end{table}

We report in \tabref{BestFit} a summary of the 1 $\sigma$ constraints on the parameters for the various fits performed.
When the lower or upper range coincides with the chosen prior, the constraint is reported in italicized characters. 
The $\chi^2$ values for each fit, also broken into the sub-data set, used are reported in \tabref{FitSummary}.
We can use the results from \tabref{BestFit} to derive the systematic uncertainties on $\delta$.
Averaging between the (main) fit and the (1GV) fit we get a value of $\delta=0.30^{+0.03}_{-0.02}$.
From the fits without convection we see that $\delta$ can be upshifted by up to a value of 0.1,
while in the (diMauro) fits $\delta$ can be downshifted by a value of 0.04. We thus quote these last two
numbers as systematic uncertainties so that $\delta=0.30^{+0.03}_{-0.02}(stat)^{+0.10}_{-0.04}(sys)$.

\begin{table}[H]
  \caption{Summary of all fits.}
  \label{tab::FitSummary}
  \centering
  \begin{tabular}{l l l c c c c c c c l r r r r r r r r c r}
  \hline \hline
  & 
  & 
  \begin{turn}{90}\textbf{Data Sets}\end{turn}&
  \begin{turn}{90}$p$ VOYAGER\end{turn} &
  \begin{turn}{90}$p$ AMS-02+CREAM\end{turn} &
  \begin{turn}{90}He  VOYAGER\end{turn} &
  \begin{turn}{90}He  AMS-02+CREAM\end{turn} &
  \begin{turn}{90}${\bar{p}/p}$ AMS-02\end{turn} &
  $\qquad$ &
  \begin{turn}{90}\textbf{Number of parameters}\end{turn} &  
  $\qquad$ &
  \begin{turn}{90}\textbf{Best fit} $\bm{\chi^2}$\end{turn} &  
  \begin{turn}{90}total\end{turn} &
  \begin{turn}{90}$p$ VOYAGER\end{turn} &
  \begin{turn}{90}$p$ AMS-02+CREAM\end{turn} &
  \begin{turn}{90}He  VOYAGER\end{turn} &
  \begin{turn}{90}He  AMS-02+CREAM\end{turn} &
  \begin{turn}{90}${\bar{p}/p}$ AMS-02\end{turn} &
  $\qquad$ &
  \begin{turn}{90} \textbf{NDF}\end{turn} \\ \hline \hline
  
  (uni-PHe)     & Universal Injection&& 
                                    $\times$ & $\times$ & $\times$ & $\times$ & $      $ &&
                                    9 &&&
                                    $53.1  $ & $ 0.5  $ & $ 11.0 $ & $ 0.1  $ & $ 40.0 $ & -  $   $ &&
                                    124 \\ 
  (uni-PHePbar) & Universal Injection && 
                                    $\times$ & $\times$ & $\times$ & $\times$ & $\times$ &&
                                    9 &&&
                                    $140.4 $ & $2.4   $ & $ 34.8 $ & $ 6.4  $ & $ 78.8 $ & $  6.4 $ &&
                                    147 \\ 
  (P)           & Starting from 5$\,$GV              && 
                                    $\times$ & $\times$ & $      $ & $      $ & $      $ &&
                                    9 &&&
                                    $ 2.46  $ & $ 0.52$ & $ 1.93 $ & -$     $ & -$     $ & -$     $ &&
                                    56 \\ 
  (PHe)         & Starting from  5$\,$GV                && 
                                    $\times$ & $\times$ & $\times$ & $\times$ & $      $ &&
                                    11 &&&
                                    $26.1  $ & $ 0.4  $ & $7.5   $ & $ 0.8  $ & $ 17.3 $ & -$     $ &&
                                    122 \\ 
  \textbf{(main)}& \textbf{Starting from 5$\,$GV}    && 
                                    $\times$ & $\times$ & $\times$ & $\times$ & $\times$ &&
                                    11 &&&
                                    $ 39.0 $ & $  0.6 $ & $ 8.0  $ & $  0.4 $ & $ 22.0 $ & $7.9   $ &&
                                    145 \\
  (diMauro)     & Starting from  5$\,$GV                && 
                                    $\times$ & $\times$ & $\times$ & $\times$ & $\times$ &&
                                    11 &&&
                                    $38.4  $ & $ 0.5  $ & $6.4   $ & $  1.5  $ & $ 23.2 $ & $ 6.4 $ &&
                                    145 \\                        
  (1GV)          & Starting from 1$\,$GV    && 
                                    $\times$ & $\times$ & $\times$ & $\times$ & $\times$ &&
                                    11 &&&
                                    $ 70.3 $ & $ 2.5  $ & $ 17.3 $ & $ 4.0  $ & $ 27.4 $ & $ 19.1 $ &&
                                    175 \\ 
  (noVc-1GV)    & No convection, starting from 1$\,$GV    && 
                                    $\times$ & $\times$ & $\times$ & $\times$ & $\times$ &&
                                    10 &&&
                                    $ 82.2 $ & $ 6.4  $ & $ 11.2 $ & $ 6.4  $ & $ 46.0  $ & $ 17.2$ &&
                                    176 \\ 
  (noVc-5GV)    & No convection, starting from 5$\,$GV  && 
                                    $\times$ & $\times$ & $\times$ & $\times$ & $\times$ &&
                                    10 &&&
                                    $ 48.7 $ & $  0.6 $ & $ 5.9  $ & $  1.6 $ & $  31.5$ & $ 8.7  $ &&
                                    146 \\ \hline \hline
  \end{tabular}
\end{table}

\section{\label{sec::conclusion}Summary and Conclusion}

We have presented new constraints on the propagation of Galactic CRs 
from an (up to) 11-dimensional parameter fit to the latest AMS-02 spectra for $p$, He, and $\bar{p}/p$. 
Solar modulation is treated  within the force-field approximation, but 
the modulation potential is constrained with a novel approach,
fitting the unmodulated CR $p$ and He spectra to
recently available low-energy data from VOYAGER,
collected after the probe left the heliosphere and thus sampling the local interstellar CR flux. 
The VOYAGER data and the unmodulated spectra are fitted jointly to the AMS-02 data and the modulated spectra.

As a first attempt, we try to fit the data with a universal injection spectrum  for $p$ and He. 
We find that a universal injection is possible when fitting only $p$ and He data. In this case, the
observed difference in $p$ and He slopes of about $\sim 0.1$ can be explained
by a significant production of secondary $p$ so that the total primary plus secondary $p$ spectrum
is steepened by the required 0.1 value in the slope. 
However, this requires a quite low value of the spectral index of diffusion $\delta\sim0.15$, and implies a large production
of $\bar{p}$ which significantly overpredicts the observations. This scenario is, thus,  in the final instance, not viable.
For the main results we thus perform a fit leaving individual spectral freedom to $p$ and He.
With this additional freedom a good fit to $p$, He, $\bar{p}/p$ spectra is achieved. 
The main result is  a tight constraint on $\delta=0.30^{+0.03}_{-0.02}(stat)^{+0.10}_{-0.04}(sys)$,
where the error is dominated by systematic uncertainties rather than statistical ones.
The robustness of this result has been cross-checked against various  factors, 
like the uncertainties in the solar modulation, the choice of the diffusion model framework, 
i.e., if convection is allowed or not, and the systematic uncertainties in the 
$\bar{p}$ production cross section.
Since solar modulation is most important at low energies, 
its effect was studied using different cuts (1 and 5$\,$GV) on the  AMS-02 data.
The effect of uncertainties in the $\bar{p}$ production cross section
was, instead, tested comparing the results of the fit when different available 
determinations of the cross section are used.
Both of these effects have an order $10-20\%$ impact on the value of $\delta$,
while the most important effect is the inclusion of convection in the model,
which shifts the value of $\delta$ from $\sim 0.3$ to $\sim 0.4$.

For the other propagation parameters the results are less definitive.
The height of the Galactic halo and the normalization of diffusion present a 
well-known degeneracy, which,  not surprisingly, cannot be resolved. In this respect,
more precise ``CR-clocks'' measurements, like the ratio $^9$Be/$^{10}$Be, which will 
be available in the future from AMS-02, are necessary. 
Regarding convection and reacceleration, 
the fit above 5$\,$GV prefers large convection velocities of $v_{0,c} \agt 50\,$km/s and Alfven velocities of $v_\mathrm{A} \alt 25\,$km/s,
with large parameter errors coming from a degeneracy between convection and reacceleration.
The fit with data down to 1$\,$GV breaks this degeneracy and gives a well definite
reacceleration of  $v_\mathrm{A}= 25\pm2\,$km/s and preference for lower values of $v_{0,c} \alt 50\,$km/s.
It remains, however, unclear how robust this determination of  $v_\mathrm{A}$ is, since it relies
on data below 5$\,$GV which are significantly affected by solar modulation.
Finally, we find that a fit without convection is nonetheless possible, providing a 
good fit to the $p$, He and $\bar{p}/p$ data, and giving a similar value of $v_\mathrm{A}$.

A comparison of these results from the constraints imposed from the AMS-02 observations
of lithium, boron, and carbon will be presented in a forthcoming companion paper.

\section*{\label{sec::acknowledgments}Acknowledgments}

We wish to thank Pasquale Serpico for  numerous useful discussions, in particular on antiproton cross sections.
We thank Leila Ali Cavasonza, Jan Heisig, Michael Kr\"amer, Julien Lesgourgues and Andy Strong for helpful discussions and comments.

\bibliography{bibliography}{}
\bibliographystyle{myUnsrt}

\end{document}